\documentclass{aa}
\usepackage{txfonts}

\input epsf
\usepackage{graphicx}
\usepackage{natbib}

\begin{document}
     
\sloppypar
     
%   \thesaurus{08     % A&A Section 6: Form. struct. and evolut. of stars
%              (02.01.2;  % Accretion, accretion disks
%               02.09.1;  % Instabilities
%              08.02.3;   % Stars:binaries:general
%              08.06.3;   % Stars:classification
%              08.14.1;   % Stars: neutron
%               13.25.3;  % X-rays: general
%               13.25.5)} % X-rays: stars
%

\title{High mass X-ray binaries in the LMC: dependence on the stellar
  population age and the ``propeller'' effect}

   \author{P.Shtykovskiy \inst{1} \and M.Gilfanov \inst{2,1}}

   \offprints{pav\_sht@hea.iki.rssi.ru}

   \institute{Space Research Institute, Russian Academy of Sciences,
              Profsoyuznaya 84/32, 117997 Moscow, Russia
        \and
              Max-Planck-Institute f\"ur Astrophysik,
              Karl-Schwarzschild-Str. 1, D-85740 Garching bei M\"unchen,
              Germany,
            }

        \authorrunning{}
        \titlerunning{High Mass X-ray Binaries in the LMC}

\date{Received 12 April 2004 / Accepted 18 October 2004}

  \abstract{
We study the population of compact X-ray sources in the Large Magellanic
Cloud using the archival data of the XMM-Newton observatory. The total
area of the survey is $\approx 3.8$ square degrees with a limiting
sensitivity of $\approx 10^{-14}$ erg/s/cm$^2$, corresponding to a
luminosity of $\approx 3\cdot 10^{33}$ erg/s at the LMC distance. 
Out of $\sim 460$ point sources detected in the 2--8 keV energy band,
the vast majority are background CXB sources, observed
through the LMC. Based on the properties of the optical and near-infrared
counterparts of the detected  sources we identified 9 likely HMXB
candidates and 19 sources, whose nature is uncertain, thus providing
lower and upper limits on the luminosity distribution of HMXBs
in the observed part of LMC. 
When considered globally,  the bright end of this distribution 
is consistent within statistical and systematic uncertainties with
extrapolation of the universal luminosity function of HMXBs. However,    
there seems to be fewer low luminosity sources, $\log(L_X)\la 35.5$,
than predicted. We consider the impact of the  ``propeller effect'' on
the HMXB luminosity distribution and show that it can
qualitatively explain the observed deficit of  low luminosity sources.

We found significant field-to-field variations in the number of
HMXBs across the LMC, which appear to be uncorrelated with the star
formation rates inferred by the FIR and $H_\alpha$ emission. We
suggest that these  variations are caused by the dependence of the
HMXB number on the age of the underlying stellar
population. Using the existence of large coeval stellar aggregates in
the LMC, we constrain the number of HMXBs  as a function of time
$\tau$ elapsed since the star formation event in the range of $\tau$
from $\sim  1-2$ Myr to $\sim 10-12$ Myr.
   }

   \maketitle

%
%________________________________________________________________

\section{Introduction.}
\label{sec:intro}

First imaging X-ray observations of the \object{Large Magellanic Cloud} with
Einstein \citep{wang91} and ROSAT  \citep{hp99}
observatories revealed a moderate population of several tens of X-ray 
sources associated with this closest neighbor of the Milky Way. 
While in the soft X-ray band a
large fraction of these sources are supernovae remnants, at 
higher energies, above $\sim$ few keV, X-ray binaries provide the dominant
contribution. For example, based on the spectral hardness of bright
ROSAT sources, \citet{kahabka} identified in the entire LMC 
$\sim 20-30$ probable X-ray binary candidates with luminosity
exceeding $L_X\ga {\rm few}\times 10^{35}$ erg/s.  

The relative numbers of high and low mass X-ray binaries (H or LMXBs)
are defined by the specific star formation rate  SFR/$M_*$ of the
host galaxy \citep{grimm03,gilfanov04}.  
Owing to rather low mass, $M_*\sim2\cdot10^{9}$ M$_{\sun}$
(section \ref{sec:lmxb}), and moderate  star formation rate (SFR) of $\sim 
0.5~M_{\sun}$ yr$^{-1}$ (section \ref{sec:sfr}), this quantity 
is rather high for the LMC, with a SFR/$M_*\sim (2-3)\cdot 10^{-10}
{\rm ~yr}^{-1}$,  exceeding by a factor of $\sim 5-10$ that of our
Galaxy.  
Correspondingly, the ratio of the expected numbers of bright,
$\log(L_X)\ga 35$, X-ray binaries in the LMC equals 
$N_{LMXB}:N_{HMXB}\sim 1:7$ \citep[][]{gilfanov04}, i.e. is nearly
opposite  to that observed in the Milky Way \citep{grimm02}.  
The absolute number of HMXBs in the LMC is expected to be 
$\sim 1/4$ of the number observed in our Galaxy, corresponding to the
ratio of star formation rates in these two galaxies.

As has been shown by \citet{grimm03}, the X-ray luminosity function
(XLF) of HMXBs obeys, to the first approximation, a universal power
law distribution with a differential slope of $\approx 1.6$,
whose normalization is proportional to the star formation rate of the
host galaxy. The validity of this universal HMXB XLF has been established
in a broad range of the star formation rates and regimes and in the
luminosity range   $\log(L_X)\ga 35.5-36$. Based on the ASCA
observations of the Small Magellanic Cloud and on the behavior of the
integrated X-ray luminosity of distant galaxies located at redshifts
$z\sim 0.3-1.3$ observed by Chandra in the Hubble Deep Field North,
\citet{grimm03} tentatively suggested that the HMXB XLF is not
dramatically affected by metallicity variations.

Study of the population of high mass
X-ray binaries in the LMC is of importance for several reasons:
\begin{enumerate}
\item
The Magellanic Clouds are known to have a significant
under-abundance of metals \citep{mc_book97}. For the interstellar medium
and young stellar population in the LMC the metal abundance is
approximately $\sim 1/3$ of the solar value \citep{mc_book97,
garnett1999, korn2002}. The effects of the metallicity
variations on the population of X-ray binaries are poorly understood. 
Study of the LMC, a galaxy with a relatively well-known chemical composition
gives a unique opportunity to investigate such effects
observationally. 
\item
Owing to the proximity of the LMC, the weakest sources become reachable
with a moderate observing time.
Indeed, the sensitivity of a typical Chandra or XMM-Newton observation,
$\sim 10^{-14}$ erg/s/cm$^2$, corresponds to the luminosity of 
$\sim 3\cdot 10^{33}$ erg/s at the LMC distance. 
This opens the possibility to study the low luminosity end of the
HMXB XLF, extending by $\sim 2-3$ orders of magnitude towards low
luminosities the luminosity range studied by \citet{grimm03}.
Among other applications, this will allow study of the impact of 
propeller effect \citep{propeller} on the luminosity distribution
of HMXBs. 
\item
The proximity and moderate inclination angle of the Magellanic Clouds
(especially that of the LMC) allows one to construct Herzsprung-Russel
diagrams and accurately determine ages,  star formation
history and initial mass function of various components of the stellar
population. In particular, large aggregates of young and coeval stellar
populations of ages ranging from $\la 1-2$ Myr to $\ga 50-100$ Myr
were discovered in the LMC. The most extreme examples of these are the LMC
4 supergiant shell \citep{braun97} and R136 (\object{HD 38268}) 
stellar cluster near the center of the nebula 30 Dor \citep{massey98}.
Combined with HMXB number counts in the optically studied regions,
these results provide a unique possibility to directly determine the
efficiency of HMXB formation as a function of time elapsed since the
star formation event.   

\end{enumerate}

The proximity of the LMC, on the other hand, creates several difficulties
in its studies. Due to its large angular extent on the sky, 
$\sim 10\degr\times 10\degr$, multiple mosaiced observations are
required to achieve a meaningful coverage of the galaxy.
The low surface density of X-ray binaries results in a large fraction of  
foreground stars and background AGNs among the detected sources, thus
creating the problem of separating true galactic members from
interlopers. This was one of the main obstacles in the earlier studies
of  X-ray binaries in the LMC with moderate angular resolution,
insufficient for reliable identification of X-ray sources with optical
catalogs. With the advent of Chandra and XMM-Newton observatories, the
latter problem is to a large extent alleviated.

\begin{table*}
\renewcommand{\arraystretch}{1.2}
\caption{List of XMM-Newton observations used for analysis}
\label{tb:pnt}
\begin{centering}
\begin{tabular}{lcccccc}
\hline
Obs. ID   & Target & R.A.  & Dec.  & Instrument & Exposure & SFR$^{(1)}$\\
          &        & J2000 & J2000 &            & ksec & $M_{\sun}$/yr\\
\hline
0113000501& 0519-69.0      &05 19 23 & -69 00 52 & MOS1+MOS2  & 25+47&$4.13\cdot10^{-3}$\\
0111130201& 2E 0509.5--6734&05 09 16 & -67 31 41 & MOS1+MOS2 & 34+34&$7.77\cdot10^{-4}$\\
0062340101& 2E 0453.8--6834&04 53 41 & -68 31 07 & MOS1+MOS2 & 16+16&$1.49\cdot10^{-3}$\\
0111130301& 2E 0525.3--6601&05 25 13 & -65 59 38 & MOS1+MOS2 & 9.5+9.5&$1.83\cdot10^{-3}$\\
0134520701& AB Dor         &05 28 35 & -65 28 30 & MOS1+MOS2 & 48+48&$2.92\cdot10^{-4}$\\
0123510101& CAL 83         &05 43 33 & -68 24 07 & MOS1+MOS2 & 10+10&$1.37\cdot10^{-3}$\\
0104060101& LMC Deep field &05 31 17 & -65 55 58 & MOS1+MOS2 & 38+38&$6.35\cdot10^{-4}$\\
0109990201& DEM L71        &05 05 49 & -67 54 11 & MOS1+MOS2 & 22+23&$1.30\cdot10^{-3}$\\
0109990101& LHA 120--N63A  &05 35 51 & -66 00 32 & MOS1+MOS2 & 9.6+9.6&$1.38\cdot10^{-3}$\\
0094410101& LMC-2(f1)      &05 42 35 & -69 03 20 & PN & 8.8&$9.74\cdot10^{-3}$\\
0094410201& LMC-2(f2)      &05 42 58 & -69 28 22 & PN & 8.8&$5.51\cdot10^{-3}$\\
0094410401& LMC-2(f4)      &05 46 48 & -69 33 39 & MOS1+MOS2 & 5.2+5.2&$4.60\cdot10^{-3}$\\
0126500101& LMC X-3        &05 38 44 & -64 06 11 & MOS1+MOS2  & 20+20&$7.82\cdot10^{-5}$\\
0113000301& N103B          &05 08 41 & -68 43 53 & MOS2 & 23ks&$4.52\cdot10^{-3}$\\
0089210701& N120           &05 18 37 & -69 37 56 & MOS1+MOS2 & 36+36&$5.76\cdot10^{-3}$\\
0089210901& N206           &05 31 51 & -70 58 39 & MOS1+MOS2 & 24+24&$3.35\cdot10^{-3}$\\
0071940101& N51D           &05 26 05 & -67 27 21 & MOS1+MOS2 & 32+32&$3.37\cdot10^{-3}$\\
0127720201& Nova LMC 2000  &05 24 41 & -70 14 00 & MOS1+MOS2 & 22+22&$1.34\cdot10^{-3}$\\
0113000401& PSR 0540--69.3 &05 40 02 & -69 21 19 & MOS2 & 35&$2.29\cdot10^{-2}$\\
0113020201& PSR J0537--6909&05 37 57 & -69 08 52 & MOS1 & 37&$5.08\cdot10^{-2}$\\
0008020101& RXJ0439.8--6809&04 39 57 & -68 07 27 & MOS1+MOS2 & 15+15&$1.83\cdot10^{-4}$\\
0089210601& SNR0450-709    &04 50 26 & -70 48 48 & MOS1+MOS2 & 56+57&$6.19\cdot10^{-4}$\\
0137160201& YYMen          &04 58 12 & -75 15 01 & MOS1+MOS2 & 87+87&$3.50\cdot10^{-4}$\\
\hline
\end{tabular}\\
\end{centering}
1 -- star formation rate within the XMM-Newton FOV
determined form the FIR flux as described in sections
\ref{sec:sfr} and \ref{sec:hmxb}.
\end{table*}

In the present paper we study the population of X-ray binaries in the LMC
based on archival XMM-Newton data. 
The distance modulus of the LMC is  $m-M=18.45\pm0.10$
\citep{mc_book97}, corresponding to the distance of $D\approx 50$
kpc. The interstellar reddening varies across the LMC, with typical
values in the range $E_{B-V}\approx 0.05-0.2$ \citep{mc_book97}. A  
value of $E_{B-V}=0.075$ corresponding to the direction towards the
nominal center of the LMC is used throughout the paper.

The paper is structured as follows. XMM-Newton observations and data
analysis are described  in section ~\ref{sec:data}. 
In section ~\ref{sec:nature} we discuss the nature
of detected X-ray sources. In section
~\ref{sec:ident} we describe the HMXB search procedure and its
results. Resulting HMXB luminosity function and the  CXB
log(N)--log(S) are presented in section ~\ref{sec:hmxbcxb}. 
The impact of the propeller effect on the HMXB luminosity function is
considered in section ~\ref{sec:propeller}.
In section ~\ref{sec:age_effects} we discuss observed
spatial non-uniformity of the $N_{HMXB}/SFR$ ratio and its dependence
on the age of the underlying stellar population.  Our results are
summarized in section \ref{sec:summary}.

\begin{figure}
\includegraphics[width=0.5\textwidth,clip=true]{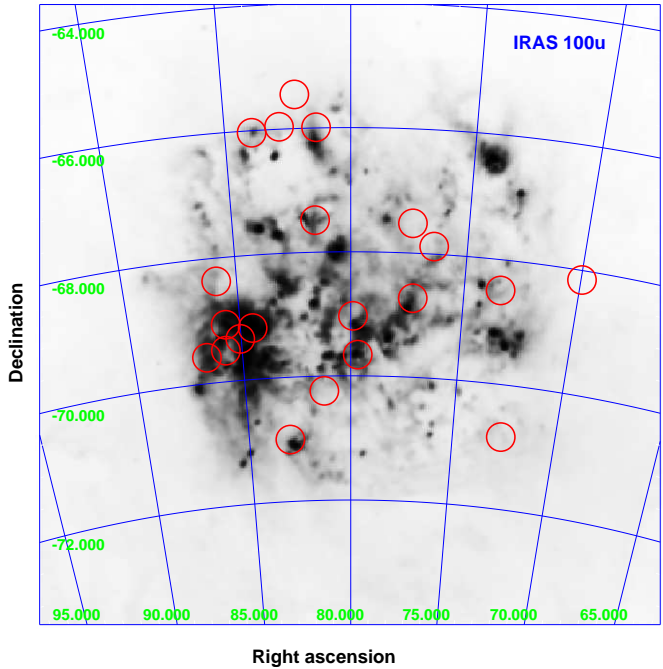}
\caption{Far-infrared (IRAS, $100\mu$) map of the $10\degr\times
10\degr$ region around the Large Magellanic Cloud. 
Circles show fields of view of the XMM--Newton observations used for
the analysis.  
}
\label{fig:fir_map}
\end{figure}

\section{Observations and data analysis.}
\label{sec:data}

We have selected 23 XMM-Newton archival observations with the pointing
direction towards the LMC and with a sensitivity better than
$\sim {\rm few}\times 10^{-14}$ erg/s/cm$^2$ in the 2--10
keV energy band. These observations are listed in Table
\ref{tb:pnt}. Fig.\ref{fig:fir_map} shows their fields of view,
overlayed on the far-infrared map (IRAS, $100\mu$) of the Large
Magellanic Cloud. 

The observations were processed with the standard SAS task
chain. After filtering out high background intervals we extracted
images in the 2--8 keV energy band. This energy range was chosen to
minimize the fraction of  supernovae remnants, cataclysmic variables
and foreground stars among detected sources, generally having softer
spectra than high mass X-ray binaries. To improve the sensitivity of the 
survey, images from MOS1 and MOS2 detectors were merged. 
If both MOS and PN data were available, we used the data having higher
sensitivity and smaller number of spurious sources.

\subsection{Source detection.}

The source detection on  the extracted images was performed with
standard SAS tasks {\em eboxdetect} and {\em emldetect}. 
The value of the threshold likelihood $L$ used in the emldetect task to
accept or reject detected source was chosen as follows.
We simulated a number of images with Poisson background counts (without
sources). Each of the generated images was analyzed with the full sequence
of source detection procedures using different values of the
emldetect threshold likelihood $L$ and for each trial value of $L$ the
number of detected ``sources'' was counted. The final value of the
threshold likelihood, $L=22$, was chosen such 
that the total number of spurious detections was $\la 3$ per 23
images. 
Note that due to the definition of the threshold likelihood in the {\em
emldetect} task, this value should not be interpreted as $L=-\ln(p)$,
with $p$ equal to the probability of detecting a given number of counts due
to the statistical fluctuation of the Poisson noise.

The obtained images  were visually inspected and  the source lists were
manually filtered of spurious 
sources near bright sources and arcs caused by a single reflection.  
All extended sources  were removed from the lists. 
The 2--8 keV source counts were converted to the 2--10
keV energy flux assuming a power law spectrum with photon index 1.7
and N$_H$=6$\cdot10^{20}$cm$^{-2}$. 
The energy conversion factors are 
ecf$_{\rm MOS}$=2.27$\cdot10^{-11}$ erg/cm$^2$ and 
ecf$_{\rm PN}$=8.0$\cdot10^{-12}$ erg/cm$^2$.
The final merged source list contains $\approx 460$ sources.
Their flux distribution is plotted in Fig.\ref{fig:lf0}.
The sources with flux $\geq10^{-13}$ erg/s/cm$^2$ are listed in Table 
\ref{tb:source}.

\subsection{Boresight correction}

For several XMM observations we performed boresight correction
using the SAS task {\em eposcorr} and optical sources from the USNO-B
catalogue \citep{usno-b}. 
The correction was applied if any of the following conditions was
satisfied:
\begin{enumerate} 
\item Optical counterparts for 2 or more well-known sufficiently
bright X-ray sources were found. 
\item Only one well-known X-ray source with an optical counterpart in
USNO is present. Offsets calculated by the {\em eposcorr} task statistically
significantly reduce its displacement from the optical counterpart.
\item The offsets calculated by the  {\em eposcorr} task 
using optical data in different non-overlapping magnitude ranges 
were consistent with each other.   
\end{enumerate}

\subsection{Correction for incompleteness}
\label{sec:area}

\begin{figure}
\includegraphics[width=0.5\textwidth]{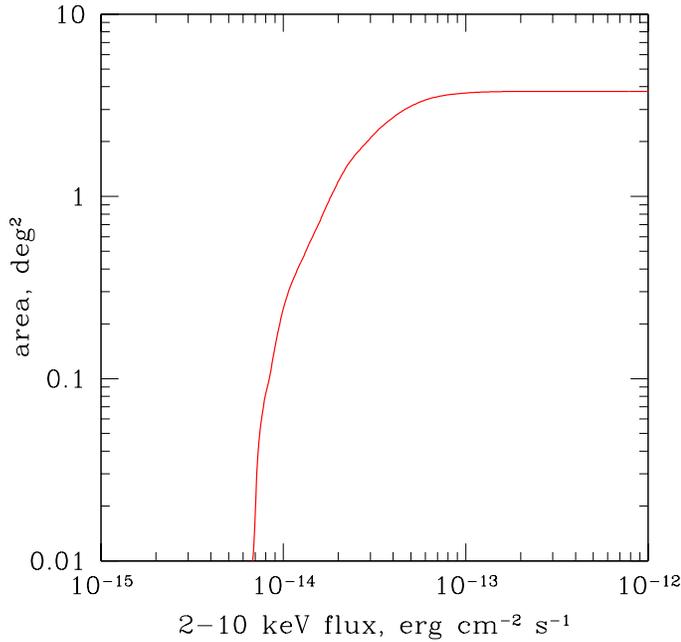}
\caption{The survey area as a function of 2--10 keV flux, calculated
as described in section \ref{sec:area}.}
\label{fig:area}
\end{figure}

To compute the flux-dependent survey area, we reproduced the
likelihood 
calculation procedure used in the {\em emldetect} task and computed
the point source detection sensitivity map for each observation. 
The sensitivity in a given pixel of the image was defined as the count
rate of the source located in this pixel, whose likelihood of
detection was equal to the threshold value used in the {\em emldetect} 
task.  
Combining sensitivity maps of all observations we calculated  
the survey area as a function of flux (Fig.~\ref{fig:area}). 
From Fig.~\ref{fig:area} one can see that the incompleteness effects
become important at fluxes  $\la10^{-13}$erg/s/cm$^2$. In the high
flux limit, the total area of the survey equals $A_{\rm tot}\approx
3.77$ deg$^2$.

The inverse area of the survey can be used to correct the observed
differential log(N)--log(S) distribution:  
\begin{equation}
\left(\frac{dN}{dS}\right)_{corrected}=\frac{1}{A(S)}\cdot\left(\frac{dN}{dS}\right)_{observed}
\label{eq:lc1}
\end{equation}
where A(S) is the survey area at flux S.
The corrected cumulative log(N)--log(S) distribution  can
be obtained as follows: 
\begin{equation}
N(>S)=\sum_{S_j>S}{\frac{1}{A(S_j)}} ,
\label{eq:lc}
\end{equation}
where $S_j$ is the flux of the j-th source. With these corrections, the flux
distributions are normalized per deg$^2$.  The flux distribution,
corrected for incompleteness effects, is plotted as the thick
histogram in Fig.\ref{fig:lf0}.

\begin{table*}
\renewcommand{\arraystretch}{1.1}
\caption{List of detected sources with 2--10 keV flux 
$F_X\geq10^{-13}$ erg/s/cm$^2$}
\label{tb:source}
\begin{centering}
\begin{tabular}{lccccl}
\hline
Source ID & $F_X$[2--10 keV] &  $L_X$[2--10 keV]$^1$ &Rmag$^{2}$& F$_x$/F$_{opt}$ &Comments       \\
        &   erg/s/cm$^2$& erg/s &   &     \\     
\hline
XMMUJ053856.7--640503 & $2.67\cdot10^{-11}$&$8.0\cdot10^{36}$& 16.7& 33 &HMXB LMC X-3   \\

XMMUJ054011.5--691953  & $2.45\cdot10^{-11}$&$7.4\cdot10^{36}$& & &Crab-like pulsar PSR B0540--69  \\
XMMUJ052844.9--652656  & $1.30\cdot10^{-11}$&&6.4$^{*}$& 0.0012  &foreground star ABDor\\
XMMUJ045818.0--751637    & $6.06\cdot10^{-12}$& &7.5$^{*}$&0.0016  &foreground star YYMen\\
XMMUJ053747.4--691020 & $5.39\cdot10^{-12}$&$1.6\cdot10^{36}$ &&&Crab-like pulsar PSRJ0537--6910\\

XMMUJ053113.1--660707 & $3.01\cdot10^{-12}$&$9.0\cdot10^{35}$ &14.0&0.31 &HMXB Be EXO053109--6609 \\
XMMUJ052402.1--701108 & $2.08\cdot10^{-12}$& && &AGN RXJ0524.0-7011 \\
XMMUJ043831.4--681203  & $8.88\cdot10^{-13}$&$2.7\cdot10^{35}$ && & \\
XMMUJ053844.2--690608     & $8.35\cdot10^{-13}$&$2.5\cdot10^{35}$ &&
& W--R stars (?) in R136 \\
XMMUJ053011.3--655123    & $6.59\cdot10^{-13}$&$2.0\cdot10^{35}$ &14.7&0.13 &HMXB Be (?) 272s pulsar$^{3}$ \\
XMMUJ050550.2--675018       & $5.80\cdot10^{-13}$&$1.7\cdot10^{35}$ &17.6 &1.5 & \\
XMMUJ053115.4--705350     & $5.25\cdot10^{-13}$&$1.6\cdot10^{35}$ &13.8 &0.043 & \\
XMMUJ053218.6--710746       & $4.28\cdot10^{-13}$&$1.3\cdot10^{35}$ &17.9 &1.6 & \\
XMMUJ053833.9--691157       & $3.65\cdot10^{-13}$&$1.1\cdot10^{35}$ &17.8&1.3\\
XMMUJ050026.6--750449    & $3.50\cdot10^{-13}$&$1.1\cdot10^{35}$ &18.2&1.7& \\
XMMUJ052847.3--653956     & $2.70\cdot10^{-13}$&$8.1\cdot10^{34}$ && &\\
XMMUJ045208.7--705652  & $2.33\cdot10^{-13}$&$7.0\cdot10^{34}$&& &  \\
XMMUJ045637.9--751015	  & $2.06\cdot10^{-13}$&$6.2\cdot10^{34}$ &18.3&1.1&\\
XMMUJ054134.7--682550     & $2.04\cdot10^{-13}$&$6.1\cdot10^{34}$ & 14.0&$0.020$  & \\
XMMUJ052312.9--701531       & $1.94\cdot10^{-13}$&$5.8\cdot10^{34}$ && & \\

XMMUJ050526.6--674313  & $1.91\cdot10^{-13}$& &10.8$^{4}$& 0.0011 &foreground star GSC 9161.1103 \\
XMMUJ053528.5--691614  &  $1.89\cdot10^{-13}$&$5.7\cdot10^{34}$ &&&SNR SN1987A  \\

XMMUJ054151.2--682609  & $1.75\cdot10^{-13}$&$5.3\cdot10^{34}$ && & \\
XMMUJ045244.2--704821   & $1.71\cdot10^{-13}$&$5.1\cdot10^{34}$ && &\\
XMMUJ050833.2--685427  & $1.70\cdot10^{-13}$&$5.1\cdot10^{34}$  &&&\\
XMMUJ052403.1--673606    & $1.67\cdot10^{-13}$&$5.0\cdot10^{34}$ &18.3&0.91& \\
XMMUJ053041.1--660535   & $1.66\cdot10^{-13}$&$5.0\cdot10^{34}$ &18.2&0.83 & \\
XMMUJ053810.0--685658   & $1.58\cdot10^{-13}$&$4.7\cdot10^{34}$ && & \\
XMMUJ052947.7--655643  &  $1.51\cdot10^{-13}$&$4.5\cdot10^{34}$  &14.8&0.033 &HMXB Be/X transient RXJ0529.8--6556  \\
XMMUJ045219.8--684151  &$1.49\cdot10^{-13}$&$4.5\cdot10^{34}$  &&& \\
XMMUJ053734.9--641012   &$1.46\cdot10^{-13}$&$4.4\cdot10^{34}$ && &  \\
XMMUJ053909.8--640310     & $1.33\cdot10^{-13}$&$4.0\cdot10^{34}$ &&&   \\
XMMUJ051832.5--693521     &$1.32\cdot10^{-13}$&$4.0\cdot10^{34}$  &&&   \\
XMMUJ051747.0--685458     &$1.32\cdot10^{-13}$&$4.0\cdot10^{34}$  &&&  \\
XMMUJ051003.9--671858 &  $1.31\cdot10^{-13}$& $3.9\cdot10^{34}$ &&&  \\
XMMUJ054237.8--683205 &$1.29\cdot10^{-13}$&$3.9\cdot10^{34}$&& &  \\
XMMUJ050736.6--684752   &$1.23\cdot10^{-13}$&$3.7\cdot10^{34}$  && &  \\
XMMUJ053257.7--705113    & $1.23\cdot10^{-13}$&$3.7\cdot10^{34}$ &19.0$^{*}$&1.3 &  \\
XMMUJ052422.4--672020  &  $1.22\cdot10^{-13}$&$3.7\cdot10^{34}$ &&&   \\
XMMUJ052353.7--672824   & $1.21\cdot10^{-13}$&$3.6\cdot10^{34}$  &&&  \\
XMMUJ054212.6--692442   & $1.19\cdot10^{-13}$&$3.6\cdot10^{34}$ & 19.3$^{*}$&1.6&  \\
XMMUJ054543.2--682528    &   $1.17\cdot10^{-13}$&$3.5\cdot10^{34}$ &&& \\
XMMUJ050048.6--750653   &    $1.15\cdot10^{-13}$&$3.5\cdot10^{34}$ &&& \\
XMMUJ053632.6--655644     &  $  1.14\cdot10^{-13}$&$3.4\cdot10^{34}$  &&& \\
XMMUJ045137.0--710210     &  $ 1.14\cdot10^{-13}$&$3.4\cdot10^{34}$  &&& \\
XMMUJ053841.7--690514      & $   1.03\cdot10^{-13}$&$3.1\cdot10^{34}$
&&& W--R stars (?) in R140\\

XMMUJ052558.0--701107  & $1.03\cdot10^{-13}$&&11.0 & 0.00068 & foreground star, K2IV-Vp, RS CVn\\

XMMUJ052952.9--705850     & $1.02\cdot10^{-13}$&$3.1\cdot10^{34}$ &&& \\
\hline
\end{tabular}\\
\end{centering}
1 -- calculated assuming distance to LMC 50 kpc.\\
2 -- apparent magnitude in the R-band of the optical counterpart from 
GSC2.2.1 catalogue \citep{gsc},  located within  3.6 arcsec from
the X-ray source.\\   
3 -- \citet{hp03}\\
4 -- apparent R-band magnitude from the USNO-B catalogue \citep{usno-b}.\\ 
\end{table*}

\section{Nature of X-ray sources in the field of LMC}
\label{sec:nature}

\subsection{Background and foreground sources}

\label{sec:cxb}

The total number of sources with the flux 
$F_X[2-10 {\rm ~kev}]>3.34\cdot 10^{-14}$ erg/s/cm$^2$ is
181. After correction for  the survey incompleteness (eq.\ref{eq:lc})
this number becomes 
$N_{\rm obs}(>3.3\cdot 10^{-14})\approx 214$. According to the CXB
$\log(N)-\log(S)$ determined by \citet{cxb}, the total number of CXB
sources expected in the field of 3.77 deg$^2$ is 
$N_{CXB}\approx 218_{-58}^{+117}$. The errors in this estimate
were computed from the uncertainty of the normalization of the  CXB 
$\log(N)-\log(S)$ relation, as given by  \citet{cxb}.
From the comparison of these numbers it is obvious that the majority of
the detected sources are background AGNs.

A significantly less important source of contamination are X-ray
sources associated with foreground stars in the Galaxy (no known
Galactic X-ray binaries were located in the field of view of XMM
observations).
There were seven well-known bright nearby stars with 2--10 keV
flux exceeding $10^{-14}$ erg/s/cm$^2$. 
All these  were excluded from the final source list.

\subsection{Low mass X-ray binaries}
\label{sec:lmxb}

Given the limiting sensitivity of the survey, $F_X\sim (1-3)\cdot
10^{-14}$ erg/s/cm$^2$, corresponding to the luminosity $L_X\sim
3\cdot 10^{33}-10^{34}$ erg/s at the LMC distance, the intrinsic LMC
sources are dominated by X-ray binaries. Their total number is
proportional to the stellar mass (LMXB) and star formation rate
(HMXB) of the galaxy. 

The total stellar mass of the LMC can be estimated from the integrated
optical luminosity. According to the RC3 catalog \citep{rc3}, the
reddening-corrected V-band magnitude  of the LMC equals 
$V_{To}\approx 0.13$, corresponding to the total
V-band luminosity of $L_V\approx 1.9\cdot 10^{9}$ $L_{\sun}$. From the
dereddened optical color $(B-V)_{To}\approx 0.44$ \citep{rc3} and using
results of \citet{m2l}, the V-band mass-to-light ratio is
$(M/L)_V\approx 0.77$ in solar units, giving the total stellar mass of the
LMC  $M_*\approx 1.5\cdot 10^9~M_{\sun}$. Using more recent
determination of the $(B-V)_{To}\approx 0.54$ \citep{lmc_ccd}, the
V-band mass-to-light ratio is $(M/L)_V\approx 1.0$, increasing the
total stellar mass estimate to $M_*\approx 2.0\cdot 10^9~M_{\sun}$.
For the stellar mass of $\approx (1.5-2.0)\cdot 10^9$ M$_{\sun}$,
using results of \citet{gilfanov04}, $\sim 8-10$ LMXBs  with
luminosity $L_X\ga 10^{35}$ erg/s are expected in the entire galaxy.

\subsection{Star formation rate in the LMC}
\label{sec:sfr}

Below we compare the estimates of the star formation rate in the LMC,
obtained from different SFR indicators.
As is commonly used, the SFR values quoted below refer to the
formation rate of stars in the $0.1-100$ M$_\odot$ range, assuming
Salpeter's initial mass function (IMF).

Based on the total $H_\alpha$ luminosity of the LMC and applying 
an extinction correction of $A(H_\alpha)\approx 0.3$
\citet{kenn95} estimated the star formation rate of 
\begin{eqnarray}
{\rm SFR}(H_\alpha)=0.26~M_\odot/{\rm yr} 
%\nonumber
\end{eqnarray}
As was discussed by \citet{kenn91}, $H_\alpha$-based SFR indicators
can underestimate the  total star formation rate in the Magellanic
Clouds, due to uncertainty of the escaping fraction of the ionizing
radiation and an unaccounted for contribution of the diffuse ionized gas.
\citet{kenn91} gave an upper limit of SFR$\la0.6~M_\odot$/yr.   

According to the Catalog of IRAS observations of large optical
galaxies \citep{rice88}, the total infra-red luminosity of the LMC is
$L_{\rm IR}\approx 2.54\cdot 10^{42}$ erg/s. With the IR--based SFR 
calibration  of \citet{kenn98}, this corresponds to the star formation 
rate of 
\begin{eqnarray}
{\rm SFR(IR)}=0.23~M_\odot/{\rm yr}
%\nonumber
\label{eq:sfr_ir}
\end{eqnarray}
i.e. it appears to be consistent with the $H_\alpha$-based
estimate. However, similarly to $H_\alpha$,  the IR luminosity
luminosity can  also underestimate the star formation rate in the
low--SFR galaxies, such as the Magellanic Clouds, due to uncertain and,
possibly, the large value of the photon escape fraction
\citep[e.g.][]{bell02}.   
An SFR estimator relatively free of this uncertainty is the one
based on combined IR and UV emission. The total integrated UV flux of
the LMC was measured by the D 2 B-Aura satellite \citep{uv80},
$F_\lambda=3.36\cdot 10^{-9}$ and $1.95\cdot 10^{-9}$ erg/s/cm$^2$/\AA~
 at $\lambda=$ 1690\AA~ and 2200\AA~ respectively.  
Correcting these value for the foreground Galactic extinction,
(E(B--V)=0.075, $A_{1690}=0.59$ and $A_{2200}=0.72$) 
and following \citet{bell02}, we obtain, with the \citet{kenn98}
calibrations, 
\begin{eqnarray}
{\rm SFR(UV)+SFR(IR)}=0.46 {\rm ~and~ }0.49 ~M_\odot/{\rm yr}
%\nonumber
\label{eq:sfr_uv_ir}
\end{eqnarray}
respectively.
These values are consistent with the SFR(UV) derived from the fluxes
corrected for both foreground and internal extinction \citep{uv80}: 
\begin{eqnarray}
{\rm SFR(UV,~ext.corr.)}=0.40-0.65~M_\odot/{yr} 
%\nonumber
\label{eq:sfr_uv}
\end{eqnarray}
where the lower and upper limits correspond to the
range of the absorption corrected fluxes in \citet{uv80}.

\citet{filipovic98}, from comparison of discrete radio (Parkes
telescope) and X-ray (ROSAT) sources, estimate a number of SNRs in
the LMC (36). From the age -- radio flux density relation
they estimate an SNR birth rate of one SNR in $100\pm20$ yrs, corresponding
to the star formation rate of:
\begin{eqnarray}
{\rm SFR(SNR~ birth~ rate)}\approx 0.7\pm0.2~M_\odot/{\rm yr}
%\nonumber
\label{eq:sfr_snr}
\end{eqnarray}

The SFR values from eq.(\ref{eq:sfr_uv_ir}--\ref{eq:sfr_snr}) seem to
be more appropriate for the physical conditions in the Magellanic
Clouds and  qualitatively agree with each other. 
In the following, we  assume: 
\begin{eqnarray}
{\rm SFR(LMC)}\approx 0.5\pm0.25~M_\odot/{\rm yr}
\label{eq:sfr_lmc}
\end{eqnarray}

The SFR value derived above can be used to predict the 
expected number of high mass X-ray binaries from HMXB--SFR calibration
of \citet{grimm03}. In applying the HMXB--SFR relation we note that
SFR values in the \citet{grimm03} sample are predominantly defined by
the FIR and radio-based estimators. The 
recent re-calibration of these SFR indicators by \citet{bell02}
restored the consistency with other SFR indicators and with the
observed normalization of IR--radio correlation, but  
resulted in the SFR-radio and SFR-IR relation being by a factor of $\sim
2.2$ and $\sim 0.7$ different from the commonly used ones
\citep{condon92,kenn98}.  With this new calibration, the SFR values in
\citet{grimm03} should effectively correspond to $\sim 1/3-1/2$ of the
total formation rate of stars in the $0.1-100~M_\odot$ mass
range. With this in mind, we change the coefficients in the equations
(6) and (7) in \citet{grimm03} to 1.1 and 1.8 respectively and use
the SFR corresponding to the total mass range, Salpeter IMF, as it is
commonly assumed.

\subsection{High mass X-ray binaries}
\label{sec:hmxb}

For the star formation rate of SFR=$0.5\pm0.25~M_{\sun}$ yr$^{-1}$, 
we predict $\sim60\pm30$ HMXBs with luminosities $\geq10^{35}$erg
s$^{-1}$ in the whole LMC. This confirms that the population of X-ray
binaries in the LMC is dominated by HMXBs. 

To predict the number of HMXBs in our sample of X-ray sources we
estimate the star 
formation rate in the part of the LMC covered by XMM pointings from IRAS
infrared maps provided by {\em SkyView} \citep{skyview}. 
Calculating the far infrared flux according to the formula
FIR$=1.26\cdot 10^{-11}(2.58S_{60\mu}+S_{100\mu})$, where FIR is flux
in erg/s/cm$^2$ and $S_\lambda$ is flux in Jy 
\citep{helou85}, and integrating the IRAS maps, we obtained the total
SFR$=0.11~M_{\sun}$/yr. This number is a factor of $\sim 2$ smaller
than eq.(\ref{eq:sfr_ir}), because it was derived from the FIR flux
instead of total infrared flux. To make it consistent
with our determination of the SFR in LMC (eq.\ref{eq:sfr_lmc}), we
simply multiply the FIR-based values  by the correction factor of
$\approx 4.5$. The thus calculated star formation rate of the part of
LMC covered by XMM observations, excluding $r<4\arcmin$ circle
centered on R136 (see section \ref{sec:30dor}), is: 
\begin{eqnarray}
{\rm SFR(XMM)}\approx 0.089\pm0.045~M_\odot/{\rm yr}
\label{eq:sfr_xmm}
\end{eqnarray}
Approximately $\sim 50\%$ of this value is due to three
observations whose fields of view included the \object{30 Doradus} giant HII
region.  

With the above value of SFR we predict about $\sim 11\pm5$ HMXBs
with luminosities 
$\geq10^{35}$ erg s$^{-1}$ in the observed part of the LMC. The error in
this number accounts for the uncertainty in the SFR estimate and does
not include the Poisson fluctuations.
We note that there already are 5 well-known high mass X-ray binaries
in our sample: LMC X--3, EXO053109--6609, RXJ0529.8--6556,
RXJ0532.5--6551 and RXJ0520.5--6932 (Table \ref{tb:source}). For comparison, the expected number of CXB sources with flux $F_X\ga 3.34\cdot 10^{-13}$
erg/s/cm$^2$ corresponding to luminosity $L_X\geq10^{35}$ erg/s,
equals $N_{\rm CXB}(F_X\ga 3.34\cdot 10^{-13})\approx 6.4$.

There are three
``historical'' bright high mass X-ray binaries in the Large Magellanic
Cloud,  LMC X-1 ($L_X\sim 1.5\cdot 10^{38}$ erg/s), 
LMC X-3 ($L_X\sim 1.5\cdot 10^{38}$ erg/s) 
and LMC X-4 ($L_X\sim 0.4\cdot 10^{38}$ erg/s). Their number is
consistent with the expected value, 
$N_{\rm HMXB}(L_X\geq 0.4\cdot 10^{38})\sim 1.5\pm 0.6$. 
Similarly, there is one bright LMXB, LMC X-2 with the average luminosity
of $L_X\sim 1.5\cdot 10^{38}$ erg/s, which is also consistent with
the expected number 
$N_{\rm LMXB}(L_X\geq 1.5\cdot 10^{38})\sim 0.23$.

\begin{figure}
\includegraphics[width=0.5\textwidth]{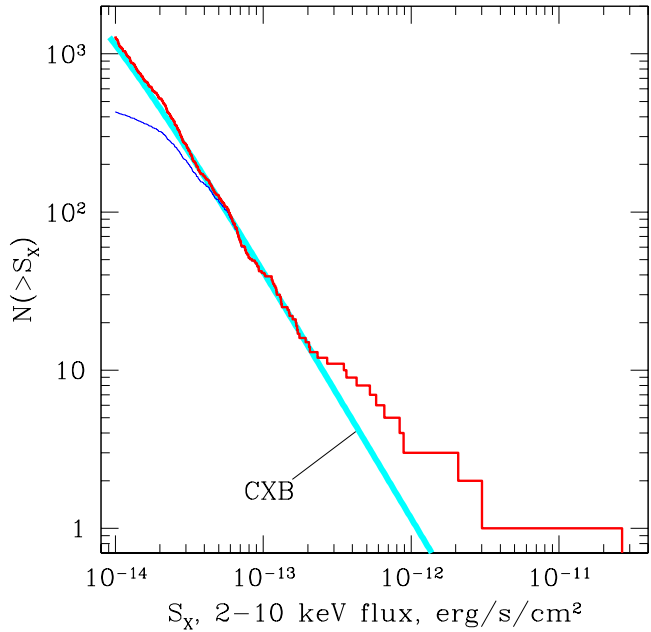}
\caption{Cumulative log(N)--log(S) distribution of detected point-like  
sources, excluding several  known foreground stars and rotation
powered pulsars (section \ref{sec:nature}). 
The thin and thick histograms show, respectively, the
observed distribution and the distribution corrected for the
incompleteness effects, as described in section \ref{sec:area}.  
The thick  grey line shows the log(N)--log(S) of CXB sources according  
to \citet{cxb}.
}
\label{fig:lf0}
\end{figure}

\subsection{Other intrinsic LMC sources}

An important type of X-ray sources associated with star formation are
Wolf-Rayet stars. Due to strong stellar wind
with $\dot{M}_{wind}\sim10^{-5}$--$10^{-4}$ M$_{\sun}$yr$^{-1}$, 
a W--R star forming a binary system with another W--R or OB star
can become a rather luminous X-ray source.  
X-ray emission in such a binary system originates from a shock formed
by colliding stellar winds of its members \citep{cherepashchuk}. 
Typical luminosities  usually do not exceed
$L_X\la$~few$\cdot10^{34}$erg s$^{-1}$.   

There are about 130 known W--R stars in the whole LMC.
We have cross correlated our data with the fourth catalogue of W--R
stars in the LMC by \citet{breysacher99}.
Four sources from our sample have been identified as W-R stars.
Two of them are faint objects with fluxes $\sim$few$\cdot10^{-14}$ 
erg cm$^{-2}$s$^{-1}$ identified in optics as WR+OB systems and
therefore were excluded from the final source list. 
The other two sources, XMMUJ053844.2--690608 and
XMMUJ053833.9--691157, located in 30 Dor, will be discussed in section
\ref{sec:30dor}.

Two well-known Crab-like pulsars located in the field
of view of XMM-Newton observations of LMC -- 
PSR B0540--6910 and PSR J0537--6910 -- were also excluded from the final
source list.

\subsection{The Log(N)--Log(S) distribution}

As demonstrated above, the population of compact X--ray sources in the
LMC field is dominated by two types of sources -- background AGNs and
high mass X-ray binaries in the LMC.  
Their log(N)--log(S) distributions  in the flux range of interest can
be described by a power law 
with the differential slopes $\approx 2.5$ (CXB) and 
$\approx 1.6$ (HMXBs). Due to a significant difference in the
slopes, their relative contributions depend strongly on the 
flux. At large fluxes, $F_X\ga 
(2-3)\cdot 10^{-13}$ erg/s/cm$^2$ ($L_X\ga 10^{35}$ erg/s) the X-ray
binaries in LMC prevail. On the contrary, in the low 
flux limit, e.g. near the sensitivity limit of our survey, $F_X\sim
10^{-14}$ erg/s/cm$^2$, the vast majority of the X-ray sources are
background AGN.

This is illustrated by Fig.~\ref{fig:lf0}, showing the observed and
corrected for incompleteness  log(N)--log(S) distribution of all
sources from the final source list. The corrected log(N)--log(S)
distribution agrees at low fluxes with that of CXB sources. At high
fluxes, there is an apparent excess of sources above the numbers
predicted by the CXB sources log(N)--log(S), due to the contribution
of HMXBs.

\section{Identification of HMXB candidates}
\label{sec:ident}

To filter out contaminating background and foreground sources, we use
the fact that optical emission from HMXBs is
dominated by the optical companion, whose properties, such as 
absolute magnitudes and intrinsic colors, are
sufficiently well known.  
The X-ray-to-optical flux ratios of HMXBs also occupy a rather
well-defined range. In addition, we take into account the fact that
intrinsic LMC objects have small proper motions, $\la 1-2$ mas/yr
\citep{mc_book97}, which helps to reject a number of foreground stars 
with high proper motion.

\subsection{Optical properties of HMXBs counterparts}
\label{sec:hmxbopt}

High mass X-ray binaries  are powered by accretion of mass
lost from the massive early-type optical companion. 
The mechanism of accretion could be connected to either (i) strong
stellar wind from an OB supergiant (or bright giant) or (ii)
an equatorial circumstellar disk around Oe or Be type star
\citep[e.g.][]{corbet, jvp95}. 
Taking into account positions of possible optical counterparts on the
Hertzsprung-Russel diagram, the distance modulus of the LMC, 
$(m-M)_0=18.45\pm0.1$ \citep{mc_book97}, and foreground and intrinsic
reddening towards the LMC, E(B--V)$\sim0.1-0.2$ \citep{mc_book97}, we can
estimate visual  magnitudes of HMXB optical companions.  
OB supergiants and bright giants (luminosity classes I--II)  have
absolute magnitudes  
M$_V\sim -7\div-4$, corresponding to an apparent magnitude in the range
m$_V\sim11.5\div14.5$. The position of Oe and Be stars in the H--R diagram
is close to the main sequence. The majority of such systems have
optical companions with a spectral class earlier than B3. Indeed, in
the catalog of high mass X-ray binaries of \citet{hmxbcat}, in only
two systems was the optical counterpart classified as B3Ve and in
$\sim 4-5$ -- later than B3, out of $\sim 90$ HMXBs with known optical
counterparts. 
Therefore the absolute magnitudes of majority of Be/X binaries are
brighter than  $M_V\la -1.6({\rm B3Ve}) \div -2.4 ({\rm B2Ve})$.  
This corresponds to a visual magnitude $m_V\la 16.0\div16.9$ and about
$\approx 0.1$ magnitude brighter in the R-band.
Accounting for the interstellar extinction, $A_V\sim 0.3-0.6$,  we
conclude that the majority of HMXBs have apparent magnitudes 
brighter than $m_V\la 16.5-17.5$, $m_R\la16.2-17.0$. 
Such magnitude filtering rejects the majority of AGN typically 
having fainter optical counterparts. 

As potential optical counterparts of HMXBs belong to spectral classes
earlier than $\sim$ B3, their intrinsic optical and near-infrared
colors are constrained by  $R-I\la-0.19$, $B-V\la-0.20$,
$V-R\la-0.08$, $J-K\la-0.16$, $I-K\la-0.34$. Taking account of the
interstellar reddening, the apparent intrinsic colors will not
exceed $\la 0.1-0.2$.

The interstellar extinction is known to vary across the LMC.
In the above estimates, the upper range of the usually quoted values
\citep{mc_book97} was used. 
A several times higher extinction is observed in the 30 Doradus
region, $E(B-V)=0.3-0.6$, with the mean value in the
central region $E(B-V)=0.44\pm0.22$, and $E(B-V)=0.65$ towards 
the center of the nebula \citep{mc_book97}. Another complication is
related to the fact that two (USNO-B and GSC) out of the three
catalogs used for initial search of optical counterparts of X-ray
sources (section \ref{sec:id_proc}) have ``holes'' with a diameter of
$\sim 10\arcmin-15\arcmin$ in the 30 Doradus region. 
This region is considered separately in section \ref{sec:30dor}.

%This mostly affects the
%magnitude threshold, whose value in the 30 Dor region should be
%decreased to $m\sim 18^m$.

The above discussion is based on our knowledge of HMXB optical
counterparts in the Milky Way. The depleted metallicity in the LMC
will, of course, affect the optical properties of the HMXB
counterparts. From  the stellar evolution studies it is known that,
for LMC metallicity, the effective temperature of the early-type stars
on the main sequence increases by $\sim 0.01-0.05$ dex
\citep{schraer93}, resulting in 
the intrinsic colors being by $\sim 0.1-0.2$ magnitude bluer than the
Galactic ones. As such, these changes do not affect the efficiency of
our selection criteria. The difference in metallicity might have 
a more significant effect on formation of HMXB systems. These effects have
not been studied yet and are one of the subjects of this paper. 
On the other hand, from the optical spectroscopy of 14 known HMXBs in
LMC, \citet{coe02} concluded that their overall optical properties are
not very different from the observed Galactic population.
 This suggests that the selection criteria based on the optical properties
of the Galactic HMXBs would be able to identify most of the HMXBs in
the LMC, with the exception of a small fraction of peculiar objects, which in
the case of the Milky Way constitute less than $\sim 5-10\%$ of the
total population of HMXBs.

\begin{figure*}
\begin{centering}\hbox{
\includegraphics[width=0.45\textwidth]{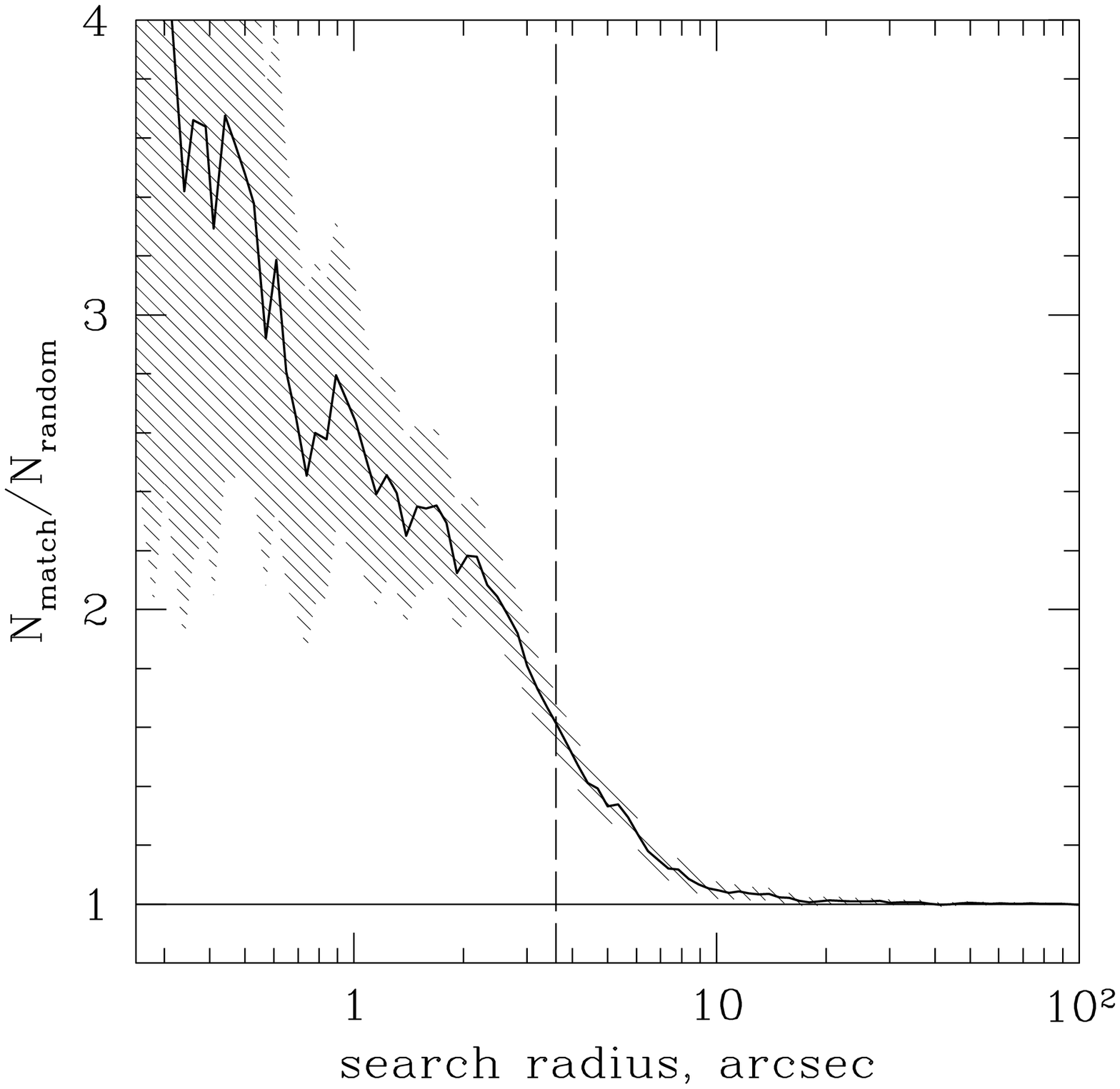}
\includegraphics[width=0.45\textwidth]{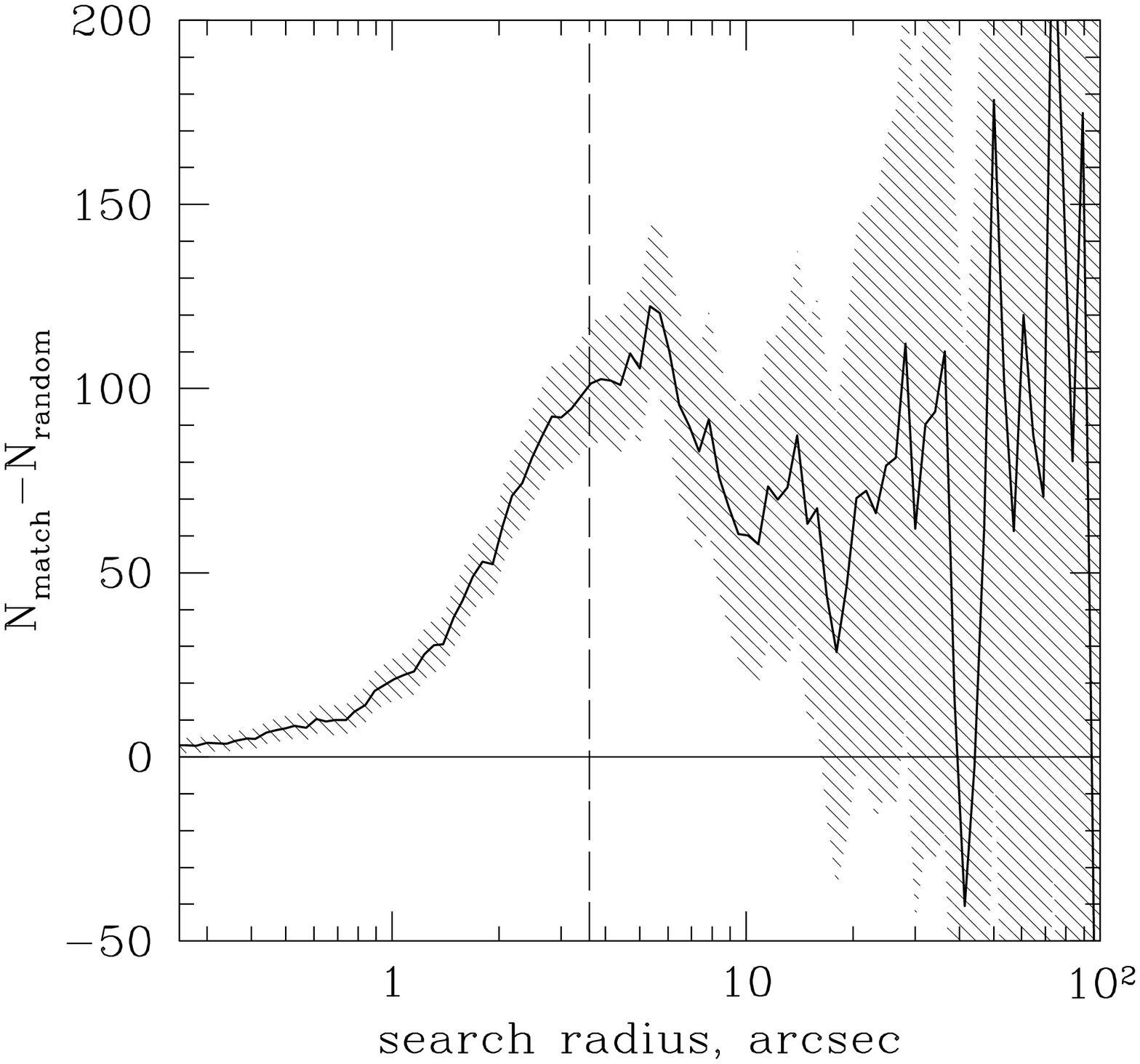}
}\end{centering}
\caption{{\em Left:} Ratio of the actual number of matches between
the X-ray source list and USNO-B catalog to the the expected number of
chance coincidences, as a function of the search radius. {\em Right:}
Difference between the number of matches and the expected number of
chance coincidences. The vertical dashed line in both panels shows the
search radius of $3.6$ arcsec. The shaded area indicates the Poisson
uncertainty, computed from the square root of the actual number of
matches.  
}
\label{fig:nmatch}
\end{figure*}

\subsection{Catalogs and selection criteria}
\label{sec:id_proc}

We have used the following optical and near-infrared catalogs:
\begin{enumerate}
\item USNO-B, version 1.0  \citep{usno-b}
\item Guide Star Catalog, version 2.2.1 (GSC2.2.1) \citep{gsc}
\item The CCD survey of the Magellanic Clouds \citep{massey}
\item 2-micron All Sky Survey (2MASS) \citep{2mass}
\item The point source catalogue towards the Magellanic Clouds
extracted from the data of the Deep Near-Infrared Survey of the Southern
Sky (DENIS) \citep{denis} 
\end{enumerate}

As a first step, we cross-correlated the XMM-Newton X-ray source list
with the USNO-B, GSC and \citet{massey} optical catalogs, using a search
radius of 3.6 arcsec and the following selection criteria: 
$R<17.5$,  $R-I<1.5$, $B-R<1.5$ and $V-R<1.5$.
If an optical object was present in the USNO-B and GSC catalogs, the
preference was given to GSC, as it has a higher photometric accuracy. 
The color limits are significantly higher than the possible colors
of HMXB optical companions and were chosen to account for the
limited photometric accuracy of optical catalogues. Relaxed color
limits also allow for variations of the intrinsic LMC reddening
up to $E(B-V)\sim$1.5,
corresponding to N$_H\sim8\cdot10^{21}$ cm$^{-2}$. 
We have found  optical counterparts for 78 X-ray sources. 
Taking into account the surface density of the optical objects in the
vicinity of X-ray sources and the value of the search radius, we
estimate the number of random matches to be 30, i.e. about half of all
optical matches. 

In the second stage, we cross-correlated all X-ray sources having
optical counterparts with the near-infrared 
catalogs 2MASS and DENIS, using the same search radius as before. 
The following filtering procedure was then applied.
\begin{enumerate}
\item 
For  X-ray sources with near-infrared counterparts, 
all those with the infrared colors  J--K or I--K $>$0.7 were excluded
from the following consideration.
\item 
All X-ray sources whose optical counterparts had detected proper
motion were rejected (all such objects had proper motion $>10$
mas/yr, significantly exceeding that of the LMC).

\item If the optical counterpart was located at a distance of more
than two positional errors of the X-ray source plus $1.5\arcsec$, the
X-ray sources was rejected. The $1.5\arcsec$ was added to allow for the
astrometric accuracy of the XMM-Newton boresight.
%In several cases  we  obtained more accurate X-ray positions
%from the   0.2--8 keV energy band, instead of hard 2--8 keV band. 

\item All X-ray sources with low X-ray-to-optical flux ratio 
$F_X/F_{\rm opt}<10^{-3}$  were rejected. The optical flux was
calculated from the  R-band magnitude using
F$_{\rm opt}=3.83\cdot10^{-6}\cdot10^{-m_{R}/2.5}$ erg/s/cm$^{2}$.
   
Such low $F_X/F_{\rm opt}<10^{-3}$ ratios are typical for foreground
stars but not for X-ray binaries. All confirmed HMXBs in our sample
have $F_X/F_{\rm opt}>10^{-2}$, except RXJ0532.5-6551 and RXJ0520.5--6932,
which have $F_X/F_{\rm opt}\approx 3.6\cdot10^{-3}$ and $\approx 4.6\cdot10^{-3}$ respectively.  

\item If an optical counterpart was found also in the \citet{massey}
catalogue, having high photometric accuracy, we used its 
optical colors for spectral classification, rejecting  X-ray sources
with counterparts of spectral type later than A0.  

\end{enumerate}

\begin{table*}
\caption{List of HMXB candidates}
\renewcommand{\arraystretch}{1.2}
\label{tb:hmxbcand}
\begin{centering}
\begin{tabular}{lcccccccl}
\hline
\#  & R.A.  & Dec.  &  $F_X$ (1) & $L_X$ (1) &$m_R$&$V-R$ & F$_x$/F$_{opt}$ &Comments       \\
 & & &    erg/s/cm$^{-2}$& erg s$^{-1}$ &   &     \\     
\hline
\multicolumn{8}{c}{\em Likely HMXB candidates$^{(3)}$}\\
\hline
%142%84.7361 & -64.0843
%05 38 56.7&-64 05 03.5
1 &  05 38 56.7&-64 05 03   &$2.67\cdot10^{-11}$&$8.0\cdot10^{36}$& 16.7&-0.2$^{5}$& 33 &LMC X-3  \\

%132%82.8046  & -66.1187
%05 31 13.1& -66 07 07.3
2 &  05 31 13.1& -66 07 07   &$3.01\cdot10^{-12}$&$9.0\cdot10^{35}$ & 14.0&0.23 &0.31&Be/X EXO053109--6609\\

%130%82.5473   &-65.8565
%05 30 11.4 & -65 51 23.4
3 &  05 30 11.4 & -65 51 23.4   &$6.59\cdot10^{-13}$&$2.0\cdot10^{35}$& 14.7&0.18 &0.13&Be/X ? 272s pulsar$^{(2)}$ \\

%152%82.8143 &-70.8972
%05 31 15.4 & -70 53 49.9
4 &  05 31 15.4 & -70 53 50  &$5.25\cdot10^{-13}$&$1.6\cdot10^{35}$& 13.6&0.1 &0.038 & \\

%121%85.3944 & -68.4306
%05 41 34.7& -68 25 50.2
5 &  05 41 34.7& -68 25 50  &$2.04\cdot10^{-13}$&$6.1\cdot10^{34}$& 14.0& -0.03 & 0.02&\\

% 127%82.4488  -65.9453
%05 29 47.7& -65 56 43.1
6 &  05 29 47.7& -65 56 43   &$1.51\cdot10^{-13}$&$4.5\cdot10^{34}$&14.6& -0.02 &0.027&Be/X transient RXJ0529.8--6556\\

% 129%83.1355 & -65.8613
%05 32 32.5& -65 51 40.7
7 &  05 32 32.5& -65 51 41  &$5.93\cdot10^{-14}$&$1.8\cdot10^{34}$&13.4&-0.36 &0.0036&OB RXJ0532.5--6551  \\

%147 %80.1224 & -69.5322
%05 20 29.4& -69 31 55.9
8 &  05 20 29.4& -69 31 56   &$4.05\cdot10^{-14}$&$1.2\cdot10^{34}$&14.1&0.03&0.0046 &RXJ0520.5--6932\\

%128%82.8260 & -66.1249
%05 31 18.2& -66 07 29.6
%17
9 & 05 31 18.2& -66 07 30  &  $3.42\cdot10^{-14}$&$1.0 \cdot10^{34}$&16.0 &-1.3$^{6}$&0.023 &\\

\hline
\multicolumn{8}{c}{\em Sources of uncertain nature$^{(4)}$}\\
\hline
%156%80.7501 & -70.3089
%05 23 00.0& -70 18 32.0
%24
10 & 05 23 00.0& -70 18 32   &
$7.93\cdot10^{-14}$&$2.4\cdot10^{34}$&17.3 &0.45&0.17 & M\\

%143%84.3209 & -64.2048
%05 37 17.0& -64 12 17.3 
%19
11 & 05 37 17.0& -64 12 17  &
$6.19\cdot10^{-14}$&$1.9\cdot10^{34}$&17.2 &&0.063& GSC, N\\ 

%140%85.3476 & -69.6093
%05 41 23.4& -69 36 33.5
%18
12 & 05 41 23.4& -69 36 34  &
$5.84\cdot10^{-14}$&$1.8\cdot10^{34}$&17.1 &&0.11 & C\\ 

%113%73.2992 & -68.6631
%04 53 11.8&-68 39 47.2
%12
13 & 04 53 11.8&-68 39 47    &
$5.71\cdot10^{-14}$&$1.7\cdot10^{34}$&16.7&0.17&0.071& C\\

%148%82.5097 & -71.0129
%05 30 02.3& -71 00 46.4
%21
14 & 05 30 02.3& -71 00 46   &
$5.3\cdot10^{-14}$&$1.6\cdot10^{34}$&14.9 &&0.013& USNO-B\\

%119%86.1429 & -68.4706
%05 44 34.3& -68 28 14.1
%16
15 & 05 44 34.3& -68 28 14  &
$4.99\cdot10^{-14}$&$1.5\cdot10^{34}$&17.4 &&0.12& USNO-B\\

%102%80.0565 & -69.1617
%05 20 13.6& -69 09 42
%9 
16 & 05 20 13.6& -69 09 42   &
$4.72\cdot10^{-14}$&$1.4\cdot10^{34}$&16.0&&0.031& M\\

%107%79.5153 &-69.0206
%05 18 03.7&-69 01 14.2
%11
17 & 05 18 03.7&-69 01 14    &
$3.51\cdot10^{-14}$&$1.1\cdot10^{34}$&17.1&&0.063& USNO-B\\

%150%83.4456 & -70.9876
%05 33 46.9& -70 59 15.4
%22
18 & 05 33 46.9& -70 59 15.4   &
$3.46\cdot10^{-14}$&$1.0\cdot10^{34}$&14.4 &0.93&0.0052 & C\\

%144%85.1816 & -64.0292
%05 40 43.6& -64 01 45.1
%20
19 & 05 40 43.6& -64 01 45  &
$3.36\cdot10^{-14}$&$1.0\cdot10^{34}$&17.1 &&0.062& C\\ 

%168%81.0638 & -70.0765
%05 24 15.3& -70 04 35.4
%28
20 & 05 24 15.3& -70 04 35   &
$2.80\cdot10^{-14}$&$8.4\cdot10^{33}$&17.0 &&0.024 & USNO-B\\

%114%81.8874 & -65.5798
%05 27 33.0& -65 34 47.3
%13
21 & 05 27 33.0& -65 34 47  &
$2.77\cdot10^{-14}$&$8.3\cdot10^{33}$&16.2&&0.022& M\\

%166%81.7583 & -70.0878
%05 27 02.0& -70 05 16.1
%27
22 & 05 27 02.0& -70 05 16   &
$2.75\cdot10^{-14}$&$8.3\cdot10^{33}$&17.4 &&0.063 & USNO-B\\

%161%80.8135 & -70.2438
%05 23 15.2& -70 14 37.
%25
23 & 05 23 15.2& -70 14 38   &
$2.35\cdot10^{-14}$&$7.1\cdot10^{33}$&16.1 &&0.018 & USNO-B\\

%155%81.7608 & -67.4577
%05 27 02.6& -67 27 27.7
%23
24 & 05 27 02.6& -67 27 28   &
$2.11\cdot10^{-14}$&$6.3\cdot10^{33}$&14.4 &&0.0032 & USNO-B, N\\

%104%80.2818 & -69.0503
%05 21 07.6& -69 03 01
%10
25 & 05 21 07.6& -69 03 01   &
$1.66\cdot10^{-14}$&$5.0\cdot10^{33}$&16.4&&0.016& USNO-B\\

%117%81.8811 & -65.3427
%05 27 31.5& -65 20 33.7
%15
26 & 05 27 31.5& -65 20 34  &
$1.58\cdot10^{-14}$&$4.7\cdot10^{33}$&17.3&&0.034& USNO-B\\

%116%81.7314 & -65.3470
%05 26 55.5& -65 20 49.2
%14
27 & 05 26 55.5& -65 20 49  &
$1.41\cdot10^{-14}$&$4.2\cdot10^{33}$&17.5&&0.037& C\\ 

%165%81.2649 & -70.1379
%05 25 03.6& -70 08 16.4
%26
28 & 05 25 03.6& -70 08 16   &
$1.37\cdot10^{-14}$&$4.1\cdot10^{34}$&17.4 &&0.033 & USNO-B\\

\hline
\multicolumn{8}{c}{\em 30 Doradus region ($r\le4\arcmin$)}\\
\hline
%xxx%
29 & 05 38 44.2 &69 06 08  & $8.35\cdot10^{-13}$&$2.5\cdot10^{35}$&  & & &
Wolf-Rayet star in R136\\
%xxx%
30 & 05 38 41.7 &69 05 14 & $1.03\cdot10^{-13}$&$3.1\cdot10^{34}$&  & &&
Wolf-Rayet star in R140\\
\hline
\end{tabular}\\
\end{centering}
\smallskip
(1) -- 2--10 keV band; 
(2) -- \citep{hp03}; 
(3) -- optical counterpart is consistent with being an HMXB;
(4) -- the optical counterpart exists, but the information is
insufficient for its reliable classification;
(5) -- B--V color from \citet{hmxbcat}; 
(6) -- B--V color; 

Comments: 
USNO-B -- counterpart in USNO-B only;
GSC -- counterpart in GSC only;
M -- multiple non-overlapping optical and/or infrared counterparts;
C -- ambiguous color indexes;
N -- non-HMXB nature is likely;
\end{table*}

\subsection{Search radius}
\label{sec:searchrad}
To choose the search radius for cross-correlation with the optical and
infrared catalogs we analyzed the dependence of the number of matches
between the X-ray source list and the USNO-B catalog as a function of the
search radius. No magnitude or color filtering was performed for this
analysis. Furthermore, for a noticeable fraction of
optical objects, the USNO-B catalog contains multiple entries. No attempt
to filter out such multiple entries was made (unlike for the final
analysis, described in sections \ref{sec:id_proc} and
\ref{sec:id_res}). Therefore the absolute numbers of the total number
of matches and of chance coincidences should not be directly compared
with the numbers in section \ref{sec:id_proc}.

At large values of the search radius,
$r_0\ga 10\arcsec-20\arcsec$, the observed number of matches
asymptotically approaches the $N_{\rm match}\propto  r_0^2$ law, expected
for the number of chance coincidences 
(Fig.\ref{fig:nmatch}, left panel).
As the localization accuracy of XMM-Newton is sufficiently high,
generally better than $\sim$few arcsec, at small values of the search
radius, the total number of matches is dominated by true optical
counterparts of X-ray sources. As is obvious from
Fig.\ref{fig:nmatch}, the chosen value of the search radius,
$r_0=3.6$ arcsec, allows us on one hand to detect a significant
fraction, $\ga 80-90\%$, of true optical counterparts (right
panel). On the other hand, it results in a reasonable fraction of
chance coincidences, $\la 50-60\%$ (Fig.\ref{fig:nmatch}, left panel).

\subsection{Identification results}
\label{sec:id_res}
\label{sec:indsrc}

After the filtering, 28 X-ray sources were left in the list of
potential HMXB candidates (Table \ref{tb:hmxbcand}). 
Of these, 9 sources have optical properties consistent with HMXB
counterparts. These sources are considered likely HMXB candidates and
are listed in the first half of the table. All known 
HMXBs located in the filed of view of the XMM-Newton observations are
among these sources. Listed in the second half are the sources whose
nature cannot be reliably established based on the available optical
data.

Comments on the individual sources:

\#9:
This source has multiple  counterparts in USNO and GSC
catalogues. One of  USNO sources has high proper motion and  
high F${_x}$/F$_{opt}\sim10^{-2}$, therefore, it is  likely  a 
chance coincidence.     
Another three GSC and one USNO sources from significantly different
epochs are located at nearly the same position and show no evidence of
proper motion. Their color indexes have large uncertainties,
but all agree with the HMXB nature of the source. It is included in
the list of likely HMXB candidates, although it is less reliable than
other sources in this group.

\#12:
This source has been previously classified by \citet{shp00} as an HMXB
candidate because of the nearby B2 supergiant. However, its improved
X-ray position obtained by XMM  deviates by $\approx20\arcsec$ from
the proposed B2 star. This  significantly  exceeds the positional
error of $\sigma_r\approx2.7\arcsec$ making the association with the
B2 supergiant improbable. The optical matches found within
$3.6\arcsec$ do not allow us to reach definite conclusion about the
nature of this source.

\#13:
The optical counterpart of this source was found in the \citet{massey}
catalogue. The color indexes B--V=0.25 and $V-R=0.17$ and proper motion of
the nearby USNO source do not allow us to draw conclusions about its
nature.

%XMMUJ053346.9--705915 (\#??).
%The optical counterpart was found only in GSC catalogue, with
%$m_R=14.4$ and color index  $V-R=0.9$. The X-ray-to-optical flux ratio
%is rather low, F${_x}$/F$_{opt} \sim5\cdot10^{-3}$, but not enough so,
%to reliably exclude it from the list of possible HMXB candidates.

\subsection{30 Doradus region}
\label{sec:30dor}

Owing to high stellar density, the central part of this luminous HII
region is not well represented in the all-sky optical catalogs,
such as USNO and GSC.

There are 5 X-ray sources in the final XMM list within the nominal
size of the nebula, $r\le7\arcmin$ \citep{mc_book97}.
Of these, 3 are located outside $r\approx 4.5$ arcmin and a search for
their optical counterparts does not present a problem.
The remaining 2 sources, XMMUJ053844.2--690608 and
XMMUJ053841.7--690514, are located at $r\le1\arcmin$ from the center
of 30 Dor and positionally coincide with the R136 and R140 stellar clusters 
respectively.  Both sources show evidence for non-zero angular
extent in XMM data. They were detected earlier by ROSAT and were classified
as high mass X-ray binaries \citep{wang95}. 
Based on Chandra data, \citet{zwart} resolved XMMUJ053844.2--690608
into one bright source and  a number of weaker sources, all positionally 
coincident with bright early type (O3f* or WN) stars in the R136
cluster. XMMUJ053841.7--690514 is located in the R140 stellar
cluster, known to 
contain at least 2 WN stars. Based on the positional coincidence with
Wolf-Rayet stars and, mainly, on the young age of the R136 cluster,
$\la 1-2$ Myr \citep{massey98}, insufficient to form compact objects, 
\citet{zwart} suggested that all these sources are colliding wind
Wolf-Rayet binaries, rather than HMXBs. However, for the two brightest
sources the X-ray-to-bolometric flux ratios, $F_X/F_{\rm bol}\sim
10^{-5}$, exceed by $\sim 1-2$ orders of magnitude the typical values
for such objects in the Galaxy and for other sources detected by
Chandra in R136 \citep{zwart}.  

Because of this uncertainty  we exclude from further
consideration the $r\le 4\arcmin$ region, centered on the R136 stellar
cluster. We note that this does not affect our main results and
conclusions.

\begin{figure}
\includegraphics[width=0.5\textwidth]{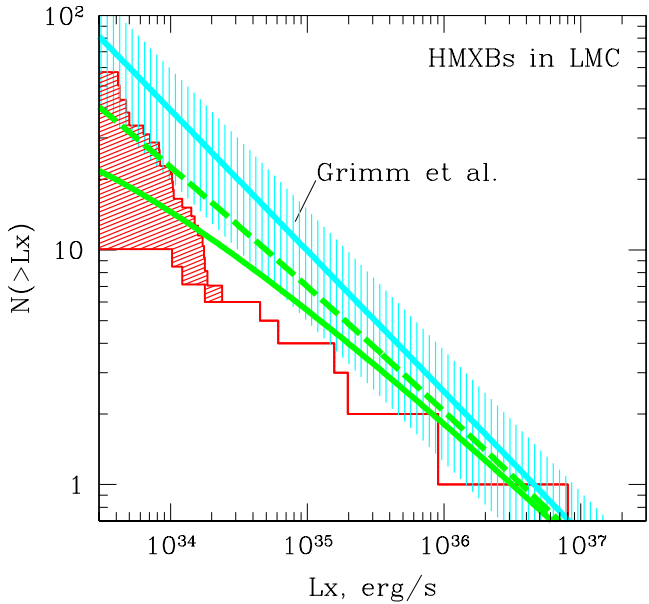}
\caption{The incompleteness-corrected  XLF of HMXB candidates in
LMC. The upper histogram shows all sources from Table
\ref{tb:hmxbcand}, except two in the 30 Dor region; the lower one
shows  likely HMXB candidates (upper part of Table
\ref{tb:hmxbcand}). These two histograms provide upper and lower
limits for the true HMXB XLF. 
The upper grey line and shaded area  show the luminosity distribution 
predicted from the ``universal'' XLF of \citet{grimm03} extrapolated
towards low luminosities and its uncertainty.   
The lower solid and dashed lines show the same XLF  modified by the
``propeller effect'' assuming the black hole fraction of 
$f_{\rm BH}=0$ and $f_{\rm BH}=0.3$ respectively (section \ref{sec:propeller}). 
}
\label{fig:lfhmxb}
\end{figure}

%RefCorrection

\subsection{Completeness}
\label{sec:optcompleteness}

The completeness of the list of HMXB candidates is defined by the
following factors: 

\begin{enumerate}
\item The completeness of the optical catalogues.

The initial search for optical counterparts is  based on the
GSC2.2 and USNO-B catalogs. The GSC2.2 is a magnitude-selected
($V\le19.5$) subset of the  GSC-II catalog 
(http://www-gsss.stsci.edu/gsc/gsc2/GSC2home.htm). The latter is 
complete to $J=21$ at high galactic latitudes \citep{gsc}. 
The USNO-B 1.0 catalog is believed to be complete down to $V=21$
\citep{usno-b}.  Completeness of both catalogs is known to
break down in the crowded regions. One example of such a region  is
the central part of 30 Dor which  was excluded from the analysis
(section \ref{sec:30dor}).  
As no sensitivity maps for the optical catalogs exist, a quantitative
estimate of the completeness of the initial counterpart search
is impossible.
However, the quoted completeness limits of both catalogs are
$2-3.5$ mag better than  chosen threshold of $17.5$ mag for the
optical counterpart search. This suggests that the completeness of
the  optical catalogs is unlikely to be the primary limiting factor.  

\item The efficiency of the initial search due to statistical and
systematic uncertainties in the positions of X-ray and optical
sources. This is probably one of the major limiting factors.
With the chosen value of the search radius, 3.6$\arcsec$, 
we detect $\sim 80-90\%$ of true matches
(section \ref{sec:searchrad}, Fig.\ref{fig:nmatch}).

\item The filtering procedure applied to the optical matches found in
the initial search. This procedure is based on expected optical and
near-infrared properties of HMXBs and, as discussed in
section \ref{sec:hmxbopt}, it would detect $\sim95\%$ of HMXBs
listed in the catalog of \citet{hmxbcat}. 
\end{enumerate}

Thus, we estimate the overall completeness of our HMXB sample to
exceed $\ga70-80\%$.
As the latter two factors are flux-dependent, they might affect 
not only the total number of detected HMXBs, but also their
luminosity distribution in the faint flux limit. 
This effect, however, is not of primary concern 
as a much larger uncertainty in the luminosity distribution at faint
fluxes is associated with the sources of unclear nature
(Fig.\ref{fig:lfhmxb}).

\section{HMXBs and CXB sources in the field of the LMC}
\label{sec:hmxbcxb}

\subsection{The luminosity function of HMXB candidates in the LMC}
\label{sec:hmxbcand}

The incompleteness-corrected luminosity distribution of HMXB
candidates is shown in Fig.~\ref{fig:lfhmxb}. The upper and lower
histograms correspond to all sources from Table \ref{tb:hmxbcand} and
to the likely HMXB candidates respectively. These two histograms
provide upper and lower limits for the true X-ray luminosity function
of HMXBs in the observed part of the LMC. As is clear from
Fig.~\ref{fig:lfhmxb}, they coincide at the luminosity $L_X\ga (2\div3)\cdot
10^{34}$ erg/s, with the uncertainty of the optical identifications
becoming significant only at lower luminosities. 
We note that the upper histogram steepens at  low luminosities
where its slope is close to that of the CXB sources.
Such behaviour indicates that a significant fraction of low luminosity 
sources of uncertain nature might be background AGNs.

In order to constrain the XLF parameters, we fit the data
using the maximum likelihood method \citep{crawford} with a power law
distribution in the range of luminosities $L_X\geq2.5\cdot 10^{34}$
erg/s, where both curves coincide.
We obtain best fit value for the differential slope
$\alpha=1.28^{+0.26}_{-0.23}$; the normalization corresponds to
$N(>10^{35}{\rm ~erg/s})\approx 5$ HMXBs. As is evident from
Fig.\ref{fig:lfhmxb}, the slope of the luminosity 
distribution appears to be somewhat flatter and its normalization 
smaller than predicted from extrapolation of the ``universal'' HMXB
luminosity function of \citet{grimm03} -- $\alpha\approx 1.6$ and 
$N(>10^{35}{\rm ~erg/s/cm^2})\approx 11\pm5$ (section \ref{sec:hmxb}).

However, the XLF flattening is not statistically  significant in the
$L_X\geq2.5\cdot 10^{34}$ erg/s luminosity range -- 
the Kolmogorov-Smirnov probability for the universal HMXB XLF model
is $\sim40\%$. Only in the entire luminosity range of
Fig.~\ref{fig:lfhmxb} is the shape of the luminosity distribution of 
reliable HMXB candidates (the lower histogram in Fig.~\ref{fig:lfhmxb})
inconsistent with the extrapolation of the universal HMXB XLF, 
the Kolmogorov-Smirnov test giving the probability of $\sim 1\%$.
On the other hand, the list of reliable HMXB candidates may be 
incomplete below $L_X\la (1-2)\cdot 10^{34}$ erg/s.

Due to the limited number of sources and ambiguity of
their optical identifications at low luminosities, it is premature to
draw a definite conclusion regarding the precise shape of the XLF below 
$L_X\la  (1-2)\cdot 10^{34}$ erg/s and its consistency with the
extrapolation to the low luminosities of the universal HMXB XLF of
\citet{grimm03}. 
We note however that the flattening of the luminosity distribution
leading to the deficit of the low luminosity sources  should be
expected due to the ``propeller effect''. As this effect and its
impact on the HMXB XLF is of interest on its own, we consider it in
detail in section \ref{sec:propeller}.   

Another factor affecting the overall normalization of the luminosity
distribution -- the dependence of the number of HMXBs on the
stellar population age -- is discussed in section
\ref{sec:age_effects}.

\begin{figure}
\hbox{
\includegraphics[width=0.5\textwidth]{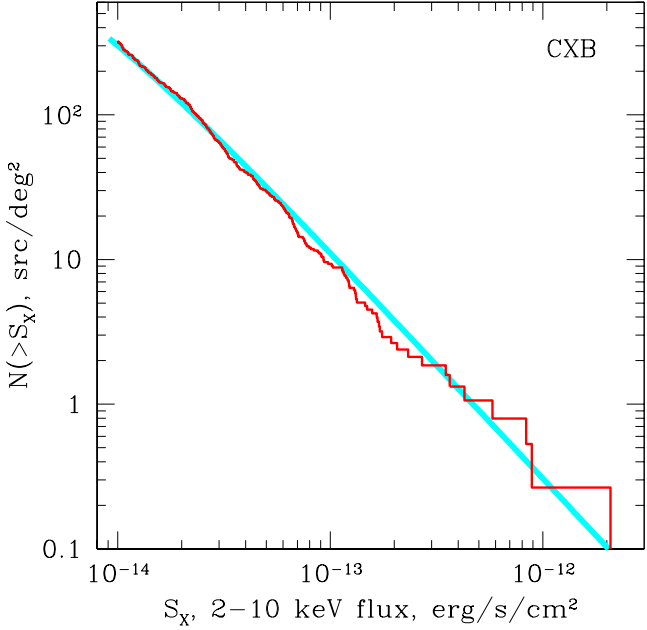}
}
\caption{Cumulative logN--logS distribution of CXB sources, obtained
after removal of  HMXB candidates. The solid line shows the
distribution from \citet{cxb}.
}
\label{fig:cxb}
\end{figure}

\subsection{Log(N)--log(S) distribution of CXB sources}
\label{sec:cxbsrc}

The log(N)--log(S) distribution of CXB sources obtained after removal
of HMXB candidates is shown in Fig.\ref{fig:cxb}. The difference
between the upper and lower limit on the HMXB population is insignificant
in this context, due to the large number of CXB sources at low fluxes.
Below we give values obtained after removing all HMXB candidates.

We fit the resulting log(N)--log(S) distribution in the flux range 
$F_X>2\cdot10^{-14}$ erg cm$^{-2}$ s$^{-1}$  with a power law model
$N(>S)=k(S/S_0)^{-\alpha}$, where $S_0=2\cdot10^{-14}$
erg/s/cm$^2$.  Best fit values  are 
$\alpha=1.62\pm 0.08$ and $k=127\pm 11$. 
The Kolmogorov-Smirnov test accepts this model, giving a K--S
probability of $\sim 16\%$. 
These best fit values agree with the CXB parameters determined in
other surveys, in particular with those from \citet{cxb}:
$\alpha=1.57^{+0.1}_{-0.08}$, k=121$^{+69}_{-31}$. 
The latter distribution is shown in Fig.\ref{fig:cxb} by the solid
line.

\section{Propeller effect and HMXB XLF} 
\label{sec:propeller}

\subsection{The ``propeller effect''}

As high mass X-ray binaries are young objects, the neutron star
magnetic field is sufficiently strong  to be dynamically important
in the vicinity of the neutron star. Indeed, the majority of known HMXBs
in the Milky Way \citep{hmxbcat} and Small Magellanic Cloud
\citep{corbet04} are X-ray pulsars. In the presently accepted picture
the transition from disk-like accretion to a magnetospheric flow,
co-rotating with the neutron star, 
occurs in a narrow region located at the magnetospheric radius
$R_m$. The  location of the transition region is defined by the
balance between the NS magnetic field pressure and the pressure (ram
and thermal) of the accreting matter. There is some uncertainty in
the definition of $R_m$, due to uncertainty in the physics of the
disk--magnetosphere interaction, the canonical value being
\citep[e.g.][]{lamb73}:    
\begin{eqnarray} 
R_m=1.4\cdot 10^9\, R_6^{10/7} M_{1.4}^{1/7} B_{12}^{4/7}\, 
L_{35}^{-2/7} {\rm ~ cm}
\label{eq:rm}
\end{eqnarray} 
where $R_6$ is the neutron star (NS) radius in units of $10^6$ cm,
$M_{1.4}$ is its mass divided by $1.4M_{\sun}$, $B_{12}$ is strength of
the magnetic filed on the NS surface in units of $10^{12}$ Gauss and
$L_{35}$  is X-ray luminosity in units of $10^{35}$ erg/s. It was
assumed above that the mass accretion rate $\dot{M}$ is 
related to the X-ray luminosity via $L_X=(GM_{NS}/R_{NS})\dot{M}$.

The character of the disk--magnetosphere interaction depends critically
on the fastness parameter, $\omega=\Omega_*/\Omega_K(R_m)$, defined as
the ratio of the neutron star spin frequency $\Omega_*$ and the Keplerian
frequency at the magnetospheric radius $\Omega_K(R_m)$.
As suggested by \citet{propeller}, at low mass accretion rates, the
spin frequency of the neutron star can exceed the Keplerian frequency
at the magnetospheric radius. In this case, corresponding to
$\omega>1$, the flow of the matter towards the neutron star 
will be inhibited by the centrifugal force exerted by the rotating 
magnetosphere and the matter can be expelled from the system due to
the ``propeller effect''. 

If the critical value of the fastness parameter $\omega_{prop}$, at
which the ``propeller effect'' occurs, is known, the corresponding
value of the critical luminosity can be computed. 
Using  eq.(\ref{eq:rm}):
\begin{eqnarray}
L_{X,prop}=3.4\cdot 10^{33}\ \omega_{prop}^{-7/3} R_6^5\, 
M_{1.4}^{-2/3} B_{12}^2\, P_{100}^{-7/3} {\rm ~ erg/s}
\label{eq:lxprop}
\end{eqnarray}
where $P_{100}$ is the NS spin period in units of 100 sec. 
The value of $L_{X,prop}$ defines the  lower limit on the
possible X-ray luminosity of an X-ray binary with 
given parameters of the neutron star. At lower values of $\dot{M}$
the system will not be observable as an X-ray source. The details of
the disk--magnetosphere interaction are not well understood,
especially  at large values of the fastness parameter.   
It is usually assumed that the ``propeller effect'' occurs at
$\omega>1$. 
However, for the accreting matter to be expelled from the system,
the linear velocity of the rotating magnetosphere should at least
exceed the escape velocity, $v_{esc}=\sqrt{2} v_K$, which 
corresponds to $\omega>\sqrt{2}$. \citet{spruit93} argued that, due to
gas pressure  effects, accretion is still possible in a narrow range
$\omega\ga 1$. They suggested that the  critical value of
the fastness parameter is $\omega\ga \sqrt{v_K/c_s}$, where $c_s$ is
the  speed of sound at the magnetospheric radius. For a standard disk
solution \citep{ss73}, this value can be very high, $\omega>>1$. 
On the other hand, the standard disk solution might be
inapplicable in the vicinity of the magnetospheric radius
\citep[e.g.][]{spruit93}. In addition, the disk might be rather
hot at the inner edge, due to dissipation
of the shock, induced by the rotating magnetosphere
\citep{propeller}.  
Owing to the strong dependence of the luminosity
on the fastness parameter (eq.(\ref{eq:lxprop})), 
this uncertainty in the critical value of  
$\omega$ translates into a much greater uncertainty in the
value of the critical luminosity. 
With this in mind, we study {\em qualitatively} the impact of the propeller
effect on the luminosity distribution of X-ray binaries with the
neutron star primary, assuming that the propeller regime occurs at 
$\omega>\sqrt{2}$.

\begin{figure*}
\includegraphics[width=0.5\textwidth]{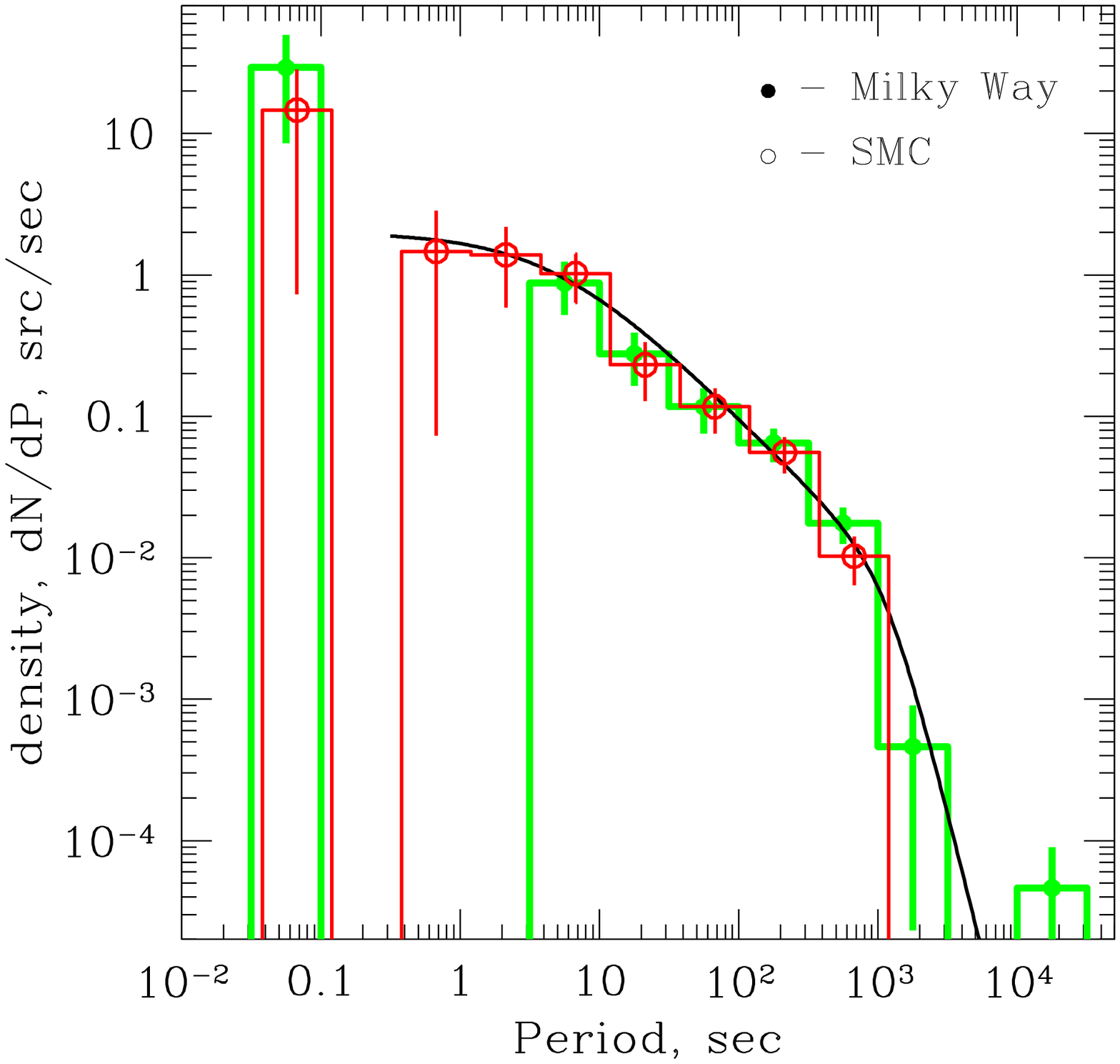}
\includegraphics[width=0.5\textwidth]{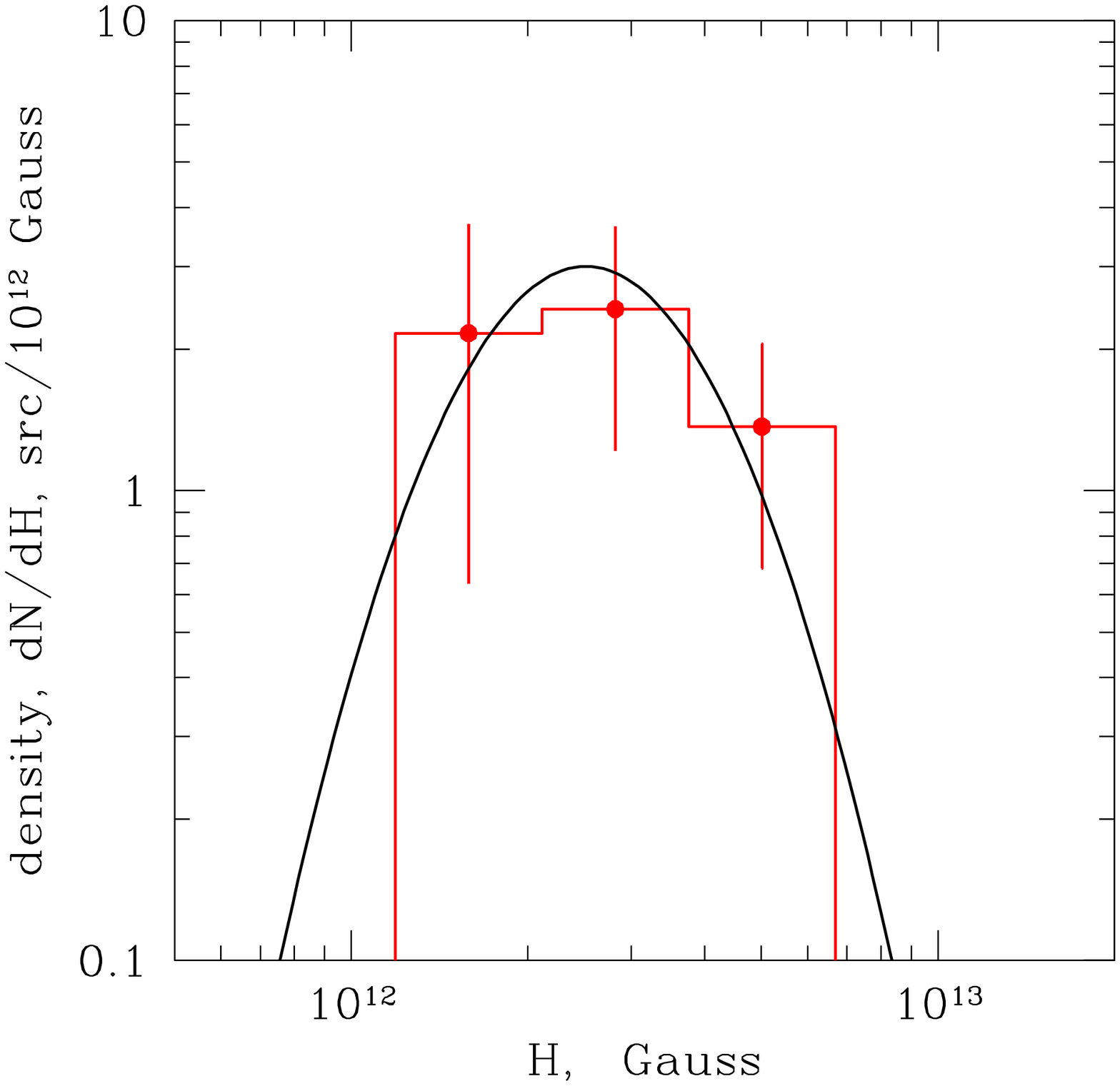}
\caption{Distribution of spin periods (left) and magnetic field
strength (right, \citet{magfield}) in high mass X-ray binaries. The spin period
distributions are shown separately for the Milky Way (
\citet{hmxbcat}) and SMC (\citet{corbet04}) sources. The SMC
distribution is shifted along the x axis by 0.2 dex for clarity.   
The smooth solid lines in both panels show the model 
distributions used for calculation of the propeller effect.}  
\label{fig:distributions}
\end{figure*}

\subsection{Impact on the HMXB luminosity distribution}

The existence of the lower limit on the luminosity of an accreting neutron
star will obviously modify the shape of the luminosity function of 
high mass X-ray binaries, leading to a deficit of low luminosity
sources.  The modified luminosity distribution can be calculated as:
\begin{eqnarray}
\frac{dN}{dL}=\left( \frac{dN}{d\dot{M}}\right)\, 
\left(\frac{R_{NS}}{GM_{NS}}\right) f(L)=
\left(\frac{dN}{dL}\right)_0 f(L)
\label{eq:xlf_prop}
\end{eqnarray}
where the luminosity-dependent factor $f(L)$ accounts for the
``propeller effect''. The undisturbed luminosity function
$\left(dN/dL\right)_0$ is defined by the the distribution of the
binary systems over the mass transfer rate $\dot{M}$ and depends on
the distributions of the binary system parameters and
the parameters of the optical companions in HMXBs. 
The function $f(L)$ is given by: 
\begin{eqnarray}
f(L)=\int\!\!\int\, \frac{dn}{dP}\, \frac{dn}{dB}\, 
\Theta(L-L_{prop}(P,B))\, dP dB
%C(L)=\int\!\!\int dP dB\, p_P(P)\, p_B(B)\, \Theta(L-L_{prop}(P,B))
\label{eq:xlf_prop1}
\end{eqnarray}
where $\Theta(x)$ is the Heavyside step function and $dn/dP$ and
$dn/dB$ are distributions of the HMXBs over the NS spin periods and
surface magnetic fields, normalized to unity. The minimal luminosity
of HMXB $L_{prop}(P,B)$ is given by eq.(\ref{eq:lxprop}), with
$\omega=\sqrt{2}$ and assuming for simplicity that all neutron stars
have the same mass and radius.   

For the undisturbed distribution $\left(dN/dL\right)_0$  
we extrapolated towards low luminosities the universal luminosity
function of HMXBs, derived by \citet{grimm03}:
\begin{eqnarray}
\left(\frac{dN}{dL}\right)_0=A\,L^{-1.6}
\label{eq:xlf0}
\end{eqnarray}
As it is mainly  based on the luminosity distribution of high
luminosity systems, $\log(L_X)\sim 36-40.5$,  it should be
relatively unaffected by the ``propeller effect'' and provides a 
reasonable approximation of the $\dot{M}$ distribution in HMXB
systems.

To estimate the distributions of HMXBs over the NS spin period and
magnetic field, we used the data on the known HMXBs. Strictly
speaking, the observed distributions are themselves modified by the
``propeller effect'', as the critical luminosity  depends strongly on
the NS period and magnetic field. For the purpose of this qualitative
consideration we ignore this effect and use observed distributions.
For the spin frequency we used the measured periods of known X-ray
pulsars  in the Milky Way \citep{hmxbcat} and Small Magellanic Cloud
\citep{corbet04}. The  distributions are plotted in the left
panel of Fig.\ref{fig:distributions}, demonstrating that 
they are similar. We approximated 
these distributions with an empirical function 
\begin{eqnarray}
\frac{dn}{dP}=\frac{n_0}{1+P/5+(P/300)^4}~~~~~~~P_{\rm min}<P<P_{\rm max}
\label{eq:distrp}
\end{eqnarray}
where P is the spin period in seconds and the constant $n_0$ is defined to
normalize the distribution to unity.
This approximation is shown in Fig.\ref{fig:distributions} by the
solid line. 
To estimate the $dn/dB$ distribution we use the results of the
determination of the magnetic field strength from observations of the
cyclotron lines by \citet{magfield}. 
The distribution is shown in the right panel in
Fig.\ref{fig:distributions} and was approximated by the log-Gaussian:
\begin{eqnarray}
\frac{dn}{dB}=n_0 \exp\left(\frac{(\log(B)-\log(B_0))^2}{2\sigma_B^2}\right)
~~~~~~~~~~~~~~~~~~~~~~~
\label{eq:distrb}
\\
B_{\rm min}<B<B_{\rm max}
\nonumber
\end{eqnarray}
with parameters $\log(B_0)=12.4$ and $\sigma_B=0.2$.

\begin{figure}
\includegraphics[width=0.5\textwidth]{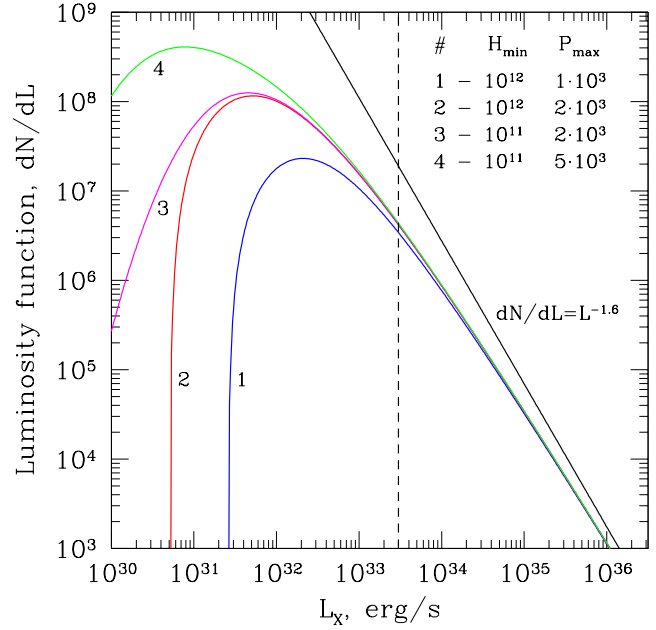}
\caption{Impact of the propeller effect on the luminosity distribution
of high mass X-ray binaries with a neutron star primary. 
The straight line shows the initial power law
distribution with differential slope 1.6. The curved lines show the
luminosity function  modified by the propeller effect for different
parameters of the distributions of the neutron star spin and magnetic
field. The neutron star mass and radius are $M_{NS}=1.4M_\odot$ and 
$R_{NS}=13$ km. 
The dashed line corresponds to our sensitivity limit.
}
\label{fig:proplum}
\end{figure}

The examples of the luminosity distributions, modified by the
``propeller effect'' for different values of the parameters are shown
in  Fig.\ref{fig:proplum}. 
Given the shapes of $dn/dP$ and $dn/dB$
distributions, the function $f(L)$ and  resulting luminosity function
weakly depend on the values of $B_{\max}\ga 10^{13}$ Gauss and $P_{\rm
min}\la 0.5$ sec. Therefore they were fixed at these values. 
The dependence on  $P_{\rm max}$ and $B_{\min}$ is significantly
stronger. The long period pulsars with $P\ga 10^3$ sec are
unaffected by the ``propeller effect'' but they can be subject to a
significant observational bias, as the long periods 
are more difficult to detect. Based on the observed period
distribution of the known X-ray pulsars, $P_{\rm max}\sim (1-2)\cdot
10^{3}$ sec seems to be a reasonable choice. 
There are no X-ray pulsars with measured $B$ below $\sim 10^{12}$
Gauss. On the other hand, for $B\sim 10^{11}-10^{12}$ Gauss, the
cyclotron line energy is in the $\sim 1-10$ keV energy range,
where it could have been easily detected by numerous experiments,
operating in the standard X-ray band. We assumed in the following 
$B_{\min}=10^{12}$ Gauss. 

Although the choice of
$P_{max}$ and $B_{min}$ significantly affects the global shape of the
luminosity function, it has a modest effect in the luminosity range of
interest, $\log(L_X)\ga 33.5$ (Fig.\ref{fig:proplum}).
In the simplified model considered above, the shape of the luminosity
function in this range is sensitive only to the radius of the neutron
star and to the critical value of the fastness parameter
$\omega_{prop}$ (eq.(\ref{eq:lxprop})).

\subsection{Comparison with the observed XLF}

The overall impact of the ``propeller effect'' on the HMXB luminosity
function depends on the fraction of neutron star binaries, as 
HMXBs with a black hole primary are, obviously,  unaffected by it. 
Presently, there is some ambiguity in this fraction. 
In the Milky Way, for 50 out of 85 HMXBs, X-ray
pulsations were detected \citep{hmxbcat}. For only a few of the
remaining 35 sources was the black hole nature of the primary
confirmed. Therefore, the fraction of HMXBs with black hole primaries
is $0.04\la f_{bh,MW} \la 0.45$. In the case of the Small Magellanic
Cloud, there are 25 objects listed in the catalog of \citet{hmxbcat},
of which for 18 X-ray pulsations were detected
\citep{hmxbcat,corbet04} and for no source was the black hole nature of
the primary confirmed, therefore  $f_{bh,SMC} \la 0.28$.

With this uncertainty in mind, we compare the model with the observed
luminosity function of HMXBs in the LMC in Fig.\ref{fig:lfhmxb}.
The luminosity distributions with account taken of the
``propeller effect'' are shown as solid and dashed thick grey lines,
computed assuming  $f_{BH}=0$ and $f_{BH}=0.3$ respectively. 
The ``propeller effect'' in both curves was computed assuming
$M_{NS}=1.4 ~M_{\sun}$, $R_{NS}=15$ km, $B_{min}=10^{12}$ Gauss, 
$B_{max}=10^{13}$ Gauss, $P_{min}=0.5$ sec, $P_{max}=10^3$ sec.
The $f_{BH}=0$ curve illustrates the maximum amplitude of the impact
of the ``propeller effect'' on the luminosity distribution of HMXBs
(for the given choice of the NS parameters and $\omega_{prop}$).

As expected, the ``propeller effect'' results in the deficit of 
low luminosity sources and  flattening of the
XLF. This behaviour is qualitatively similar to
the observed XLF (Fig.\ref{fig:lfhmxb}). 
However, due to lack of distinct features of the ``propeller effect''
in the XLF at $\log(L_X)\ga 33$ and large uncertainty in the observed
HMXB XLF at low luminosities, it is premature to draw any definite
conclusion  regarding the relevance of the ``propeller regime'' of
accretion  onto a strongly magnetized neutron star.
Its impact would be more apparent in the $\log(L_X)\la 32$
luminosity range. In order to 
construct this, a larger number of sources and better sensitivity
limits are required.

\section{HMXBs and the age of underlying stellar population}
\label{sec:age_effects}

\subsection{Spatial distribution of HMXBs in LMC}

\begin{figure}
\centerline{\hbox{
\includegraphics[width=0.45\textwidth,clip=true]{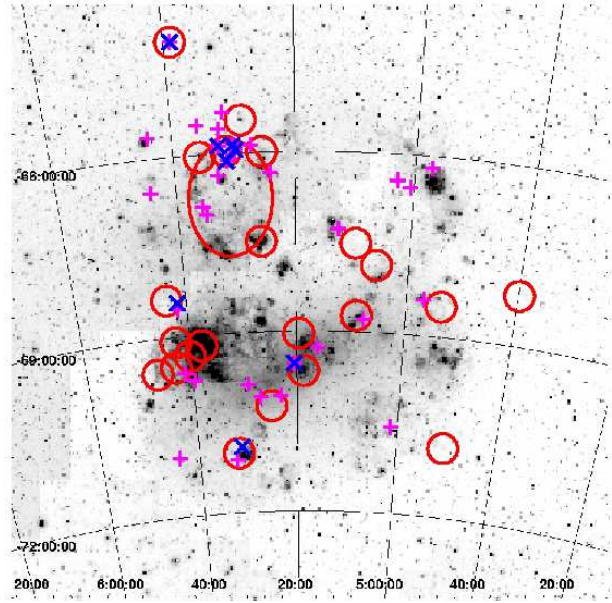}
}}
\caption{The $H_{\alpha}$ map of LMC (greyscale image). 
The overlayed symbols are: ``+''(magenta) -- all known HMXBs from the catalog
of \citet{hmxbcat} and HMXB candidates discovered by ROSAT
\citep{kahabka}; ``x'' (blue) -- likely HMXB candidates (upper part of 
Table \ref{tb:hmxbcand}). 
Large circles indicate fields of view of XMM observations.
The oval-shaped region in the north-east shows the location of the 
LMC 4 supergiant shell. The region of the highest surface brightness,
located to the south of Dec$=-69\degr$, is the 30 Doradus HII region. 
}
\label{fig:halpha_src}
\end{figure}

Considered globally, the number and luminosity distribution of HMXBs
in the observed part of the LMC roughly agree with expectation based
on the  
universal luminosity function of HMXBs, probably modified by the
propeller effect at the low luminosity end (Fig.\ref{fig:lfhmxb}). 
However, examination of the numbers of detected HMXBs in the individual
XMM pointings reveals a significant non-uniformity of the
$N_{HMXB}/SFR$ ratio, as illustrated by Fig.\ref{fig:halpha_src}.
This figure also shows all {\em known} HMXBs and HMXB
candidates in LMC. These do not represent a flux-limited
sample, therefore, it should be interpreted with caution. 
Nevertheless, their spatial distribution shows the same trends as the
distribution of the HMXBs detected by XMM, which do constitute a
flux-limited sample.
Fig.\ref{fig:halpha_src} reveals the remarkable absence
of correlation between the surface density of HMXBs and the star
formation rate, as traced by $H_\alpha$ emission. Indeed, nearly
half of the known HMXBs are located in the north-east part
of the \object{supergiant shell LMC 4} and beyond it. On the other hand, there
are only a few HMXBs in and near the 30 Dor region, which has the highest
surface density of star formation, as traced by $H_\alpha$ and FIR
emission, and accounts for $\approx 50\%$ of the total SFR in the part
of the LMC covered by the survey. 
Factors such as obscuration  and the SN kicks do affect the
spatial distribution of HMXBs, but they seem to be insufficient to
explain the observed distribution.
Indeed, the largest interstellar extinction in the LMC region
is observed towards the center of 30 Dor, $E(B-V)=0.65$
\citep{mc_book97}, corresponding to the hydrogen column density
of $N_H\approx 3.5\cdot 10^{21}$ cm$^{-2}$. This value  cannot
affect the spatial distribution of the sources
in the 2--8 keV band  in any significant way.
The typical SN kick velocities of HMXBs are unlikely to exceed $\sim
50$ km/s, which corresponds to $\sim 3.5$ arcmin/Myr at the LMC
distance. Such a velocity is obviously insufficient to
significantly modify the HMXB spatial distribution on the angular scale of
degrees  over a few million years.
A more attractive explanation is offered by the effects related to the
age of the underlying stellar population, as discussed below.

\subsection{HMXBs and the stellar age}
\label{sec:hmxb_age}

HMXBs are often considered as an ``instantaneous'' tracer of star
formation. However, the lifetime of the most massive stars of
$M\sim 125~M_\odot$  is $\sim 2.6$ Myr and  can be as long as  
$\sim 15-20$ Myr for  $M\sim 8-10~M_\odot$ stars. To this 
time one should add a delay required for the binary to reach the
HMXB phase, $\tau_{PSN}\sim 1$ Myr. 
Following an instantaneous creation of stars at $\tau=0$, the  
number of HMXBs  $N_{HMXB}(\tau)$ is a non-monotonic function  of time
$\tau$ passed  since the star formation event. It equals zero at
$\tau\approx 1-3$ Myr, unless stars are formed significantly
heavier than the conventionally accepted upper mass limit of $\sim
125~M_\odot$. At later times, the $N_{HMXB}(\tau)$ is an increasing
function of $\tau$, until at least $\tau\sim 20$ Myr, corresponding to
the lifetime of the least massive stars, $M\sim 8~M_{\sun}$, capable
of leaving behind a compact object. The behaviour of $N_{HMXB}(\tau)$
at later times is not clear. One might expect that at 
$\tau\ga 20$ Myr, the number of HMXBs will decrease with
time. However, this behaviour will be affected by the binary
systems with a less massive companion, entering the HMXB phase at later 
times (i.e. having a longer time delay $\tau_{PSN}$).

This simple picture might qualitatively explain the observed
non-uniformity of the spatial distribution of HMXBs in the LMC.
Indeed, the central region of 30 Dor has a very young age, $\approx
1-2$ Myr, as revealed by its H-R diagram and stellar mass distribution
\citep{massey98}. This is insufficient to form compact objects --
neutron stars or black holes, as was noted by \citet{zwart}, unless stars
with masses exceeding $\sim 150~M_\odot$ were formed in R136. 
The LMC4 supergiant shell, on the other hand, has an age of $\sim 
10$ Myr \citep{braun97}, which is sufficient for all stars heavier
than $\sim 20~M_{\sun}$ to have become collapsed objects.  The OB 
associations in the southern part of LMC 4 are somewhat younger, with 
ages ranging from $\sim 2$ Myr to $\sim 6-9$ Myr \citep[][and
references therein]{braun97}, which might
explain the pausity of HMXBs there. 

In the bow-shock induced star formation model of \citet{deboer98},
the stellar population age increases  clockwise from $\la 1$ Myr to the
south-east of 30 Dor, to $\sim 10-15$ Myr in the north-east of the LMC,
near LMC 4, to $\sim 20-40$  Myr to the north-west. This also
qualitatively agrees with the observed distribution of 
HMXBs in the LMC.

%\subsection{Time dependent efficiency of HMXBs formation}
\subsection{Number of HMXBs as a function of time after the star
formation event}  
\label{sec:eta_hmxb}

The proximity of LMC  allows to construct H-R
diagrams and sufficiently accurately determine the age and the mass
function of the stellar population in various parts  
of LMC. Due to the small inclination angle and, consequently, small depth,
the individual stellar associations can be studied without significant
projection effects and contamination by the foreground and background
populations.
As a result of these studies, it has been found that the supergiant
shells in the LMC often host coeval stellar populations. One of the most
intriguing examples is the LMC 4 supergiant shell, with a diameter of $\sim
1.4$ kpc, inside which no significant age gradients have been found,
with all the stars having approximately the same age of $\sim 9-12$
Myr \citep{braun97}. 
X-ray observations and HMXB number-counts in the fields, well studied
in the optical band,  open a unique possibility to directly determine
the number of high mass X-ray binaries as a function of
time elapsed since the star formation event.
This possibility is explored below using two stellar
associations in the LMC as an example -- the R136 stellar cluster and
the northern part of the LMC 4 supergiant shell.

\begin{figure}
\hbox{
\includegraphics[width=0.5\textwidth]{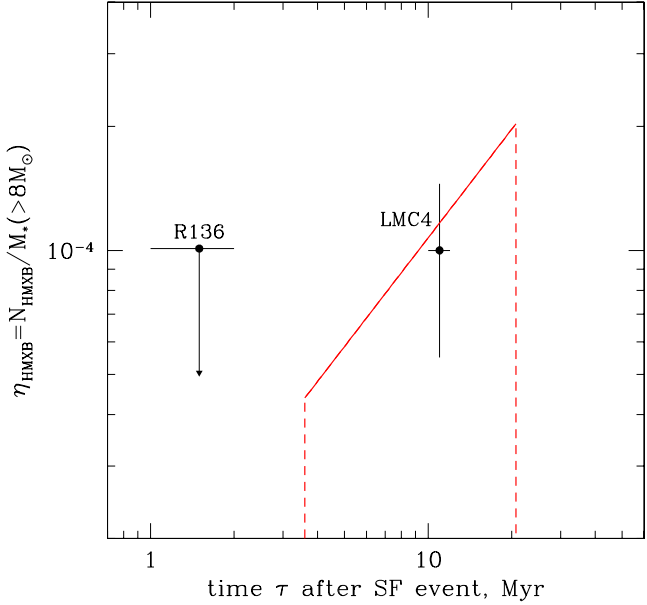}
}
\caption{The time dependence of the specific number of HMXBs,
$\eta_{HMXB}(\tau)$. The point with error bars  and the upper limit
are data on the LMC 4 supergiant shell and R136 star cluster
(Table \ref{tb:eta_hmxb}). The solid line shows  the simple model
described in section \ref{sec:eta_hmxb} with the normalization
determined from the $N_{HMXB}-$SFR relation from \citet{grimm03}.
}
\label{fig:eta_hmxb}
\end{figure}

We assume that stars  are formed instantaneously at time $\tau=0$
with the Salpeter IMF with an upper mass cut-off $M_u=125~M_{\sun}$.   
The time-dependent specific number of HMXBs, $\eta_{HMXB}(\tau)$, is
defined as the number of HMXBs present after time $\tau$
since the star formation event, normalized to the total mass of
massive stars with $M\ge 8~M_{\sun}$, formed at $\tau=0$:
\begin{eqnarray}
\eta_{HMXB}(\tau)=\frac{N_{HMXB}(\tau)}{M( >8M_{\sun}\,,\tau=0)}
\label{eq:eta}
\end{eqnarray} 
The age of the stellar population in the northern part of LMC 4 was
determined by \citet{braun97} from the analysis of the H-R diagram,
$\tau\approx 10-12$ Myr, with  little dispersion between 
individual fields.  \citet{braun97} also determined the stellar mass
function and have shown that its slope in the $\sim 3-15~M_{\sun}$
range is close to the Salpeter value of 2.35. 
The surface mass density of the initially-formed massive stars with
masses between 8 and 125 $M_{\sun}$ is $\approx 85~M_{\sun}/{\rm
arcmin}^2$. Extrapolating these  results to the XMM observation of the
LMC Deep field, covering $\approx600$ arcmin$^{2}$ and
whose FOV did not coincide with the fields studied 
by \citet{braun97} but was almost within the boundaries of LMC 4, we
can estimate the total mass of massive stars within the XMM FOV,
formed $\approx 10-12$ Myr ago, 
$M(>8M_{\sun}\,,\tau=0)\sim 5\cdot 10^4~M_{\sun}$. 
In this observation, 5 HMXB candidates with luminosity $L_X\ge 10^{34}$
erg/s were detected. With these numbers we estimate 
\begin{eqnarray}
\eta_{HMXB}(\tau=10-12 {\rm Myr})\sim 1\cdot 10^{-4}{\rm ~~HMXB}/M_{\sun}.
\label{eq:eta_lmc4}
\end{eqnarray} 
A similar estimate can be made for the R136 region.
For the R136 cluster we complimented the results of \citet{massey98} with the 
flanking fields data of \citet{sirianni00},
covering an area of $4.9$ arcmin$^2$ centered
on the R136 stellar cluster. In this region only two X-ray sources were
detected by XMM, as discussed in section \ref{sec:30dor}. As their
nature is uncertain, we assume conservatively that the total number of
HMXBs in this region is $\le2$.
The results are summarized in Table \ref{tb:eta_hmxb} and
plotted in Fig.\ref{fig:eta_hmxb}.

In a simple ad hoc model of the formation of HMXBs, the dependence
$\eta_{HMXB}(\tau)$ can be estimated as follows. 
\citet{braun97}, fitting the  stellar tracks of \citet{schraer93},
have found that the lifetime of a star of mass M is
\begin{eqnarray}
t_*\approx 86\,M^{-0.72} {\rm ~Myr}
\label{eq:lifetime}
\end{eqnarray}
where the mass of the star $M$ is expressed in solar units.
For a Salpeter mass function, the rate of type II supernovae
depends on the time $\tau$ elapsed since the star formation event as
follows:  
\begin{eqnarray}
\frac{dN_{SN}}{dt}\propto \tau^{0.88}, ~~2.6\le\tau\le 19 {\rm ~Myr}.
\label{eq:nsn}
\end{eqnarray}
In the latter inequality $\tau=2.6$ and 19 Myr are the lifetimes of
the most massive stars and of the stars with  $M=8~M_{\sun}$
respectively.
Taking into account that the HMXB lifetime $\tau_H$ is short, we obtain
for the number of HMXBs active at time $\tau$:
\begin{eqnarray}
N_{HMXB}(\tau)\propto \frac{dN_{SN}}{dt}(\tau-\tau_{PSN})\cdot
\tau_H
\label{eq:eta_hmxb1}
\end{eqnarray}
where $\tau_{PSN}$ is the post-supernova time required for the binary
to reach the HMXB phase. Here we have assumed that $\tau_{PSN}$
and $\tau_H$ are the same for all binaries. With eq.(\ref{eq:nsn}) we
obtain:
\begin{eqnarray}
\eta_{HMXB}(\tau)\propto (\tau-\tau_{PSN})^{0.88}.
\label{eq:eta_hmxb}
\end{eqnarray}
This equation is valid  for $2.6\le\tau-\tau_{PSN}\le 19$
Myr. The behaviour of $\eta_{HMXB}(\tau)$ outside this interval,
especially at larger $\tau$, is unclear, as discussed in section
\ref{sec:hmxb_age}. 
The shape of $\eta_{HMXB}(\tau)$ calculated using eq.~\ref{eq:eta_hmxb}
with $\tau_{PSN}=1$ Myr is shown in Fig.\ref{fig:eta_hmxb}. 

The normalization in equation (\ref{eq:eta_hmxb}) depends on a
number of parameters of the binary evolution, such as the total
fraction of the binaries, the fraction, survived the supernova
explosion etc, whose values are unknown.
On the other hand, normalisation can be determined from the
calibration of the $N_{HMXB}$--SFR relation obtained by \citet{grimm03}. 
From its derivation, this calibration corresponds to the case of a
steady star formation on a time scale longer than HMXB formation and
life times. Therefore:
\begin{eqnarray}
\frac{N_{HMXB}}{SFR}=\int \eta_{HMXB}(\tau) \,d\tau .
\label{eq:nhmbx_norm}
\end{eqnarray}
The normalization of the curve shown in Fig.\ref{fig:eta_hmxb} was
obtained from integration of eq.(\ref{eq:nhmbx_norm}) with 
$\eta_{HMXB}(\tau)$ defined by eq.(\ref{eq:eta_hmxb}) and
the integration limits 2.6 and 19 Myr. 
The ratio $N_{HMXB}/{\rm SFR}=N_{HMXB}(\ge 10^{34}{\rm~erg/s})/{\rm
SFR}\approx 1.9\cdot 10^3$ was calculated using eq.(7) from
\citet{grimm03} for HMXBs with $L_X>10^{34}$erg/s.
As can be seen from Fig.\ref{fig:eta_hmxb}, within the accuracy of
our (very crude) approximation, there is a good agreement with the
data.

\begin{table}
\caption{Constraints on $\eta_{HMXB}(\tau)$}
\renewcommand{\arraystretch}{1.2}
\label{tb:eta_hmxb}
\begin{centering}
\begin{tabular}{lcccc}
\hline
Name & Age, Myr &  $M_*^{(1)}$ & $N_{HMXB}^{(2)}$& $\eta_{HMXB}^{(3)}$\\
\hline
R136 & $\approx 1-2$ & $2\cdot 10^4$ & $\le 2$ & $\le 1\cdot 10^{-4}$\\
LMC 4 & $\approx 10-12$ & $5\cdot 10^4$ & $5$ & $(1\pm 0.45)\cdot 10^{-4}$\\
\hline
\end{tabular}\\
\end{centering}
\smallskip
(1) -- Total mass of massive stars, $M\ge8~M_{\sun}$;
(2) -- number of HMXBs detected by XMM; 
(3) -- number of HMXBs per unit stellar mass; References for age and
IMF: R136 -- \citet{massey98} and \citet{sirianni00}, LMC 4 --
\citet{braun97}
%, NGC1818 -- \citet{ngc1818} 
\end{table}

The above model is of course very simplified.
In its more realistic version the effects of spread in
$\tau_{PSN}$, $\tau_{HMXB}$ and in ages of the stellar population
should be taken into account. However, to a first approximation
such effects will only smooth the edges, leaving the overall behaviour
of $\eta_{HMXB}(\tau)$ unchanged. Significantly more important are
the effects of evolution of the secondary and contribution of the 
intermediate mass systems.

% Note that a similar calculation for the total luminosity results
%in about $\sim 10$ times difference between predicted and observed
%value of normlization of $l_{HMXB}(\tau)$. This disagreement can be
%easily explained by the effects of statistics of small numbers,
%discussed in \citet{stat}.

\begin{figure}
\hbox{
\includegraphics[width=0.5\textwidth]{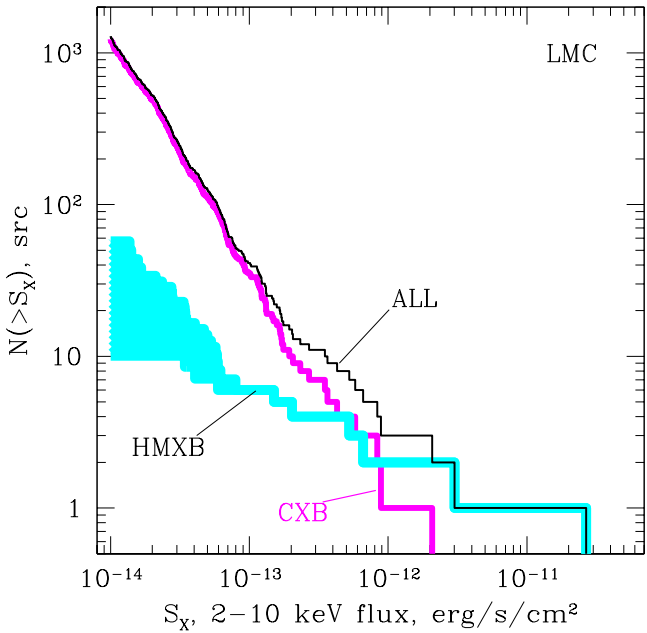}
}
\caption{The log(N)-log(S) distribution of X-ray sources in the LMC
field. The different histograms show the distribution of all sources, high
mass X-ray binaries and CXB sources. The shaded area in the
distribution of HMXBs corresponds to the range of uncertainties 
discussed in section \ref{sec:hmxbcand}.
}
\label{fig:lfxx}
\end{figure}

\section{Summary}
\label{sec:summary}

Based on the archival data of XMM--Newton observations, we studied the
population of compact sources in the field of the LMC. The total area of
the survey is $\approx 3.8$ sq.degr. with a limiting sensitivity of
$\sim 10^{-14}$ erg/s/cm$^2$ (Fig.\ref{fig:area}), corresponding to
the luminosity of $\sim 3\cdot 10^{33}$ erg/s at the LMC distance. 
\begin{enumerate}
\item
Out of the 460 compact sources detected in the 2--8 keV energy
band the vast majority, $\ga 94\%$, are CXB sources, observed through
the LMC (section \ref{sec:cxb}, Figs.\ref{fig:lf0},\ref{fig:lfxx}). 
\item
Based on the stellar mass and the star formation rate of the LMC we
demonstrate that the majority, $\ga 75-90\%$, of the intrinsic LMC
sources detected in the 2--8 keV band are high mass X-ray binaries 
(section \ref{sec:nature}). 
\item
The proximity of the LMC and the adequate angular resolution of
XMM--Newton make it possible to reliably filter out the sources
whose properties are inconsistent with being an HMXB. 
Based on the optical and infrared
magnitudes and colors of the optical counterparts of the X-ray sources
we identify 9 likely HMXB candidates (6 of which were previously
known HMXBs or HMXB candidates) and 19 sources of uncertain
nature (section \ref{sec:ident}, Table \ref{tb:hmxbcand}).  
The remaining $\sim 440$ sources, with a few exceptions, are
background objects, constituting the resolved part of CXB. Their
flux distribution is consistent with other determinations of
the  CXB log(N)--log(S) (section \ref{sec:cxbsrc},
Fig.\ref{fig:cxb}). 
\item
With these results we constrain lower and upper bounds of the
luminosity distribution of HMXBs in the observed part of the LMC.
We compare these with the extrapolation towards low luminosities of the 
universal luminosity function of HMXBs, derived by \citet{grimm03}. 
At the high luminosity end, the observed distribution is consistent
with the extrapolation of the universal XLF. The number of bright,
$L_X\ge 10^{38}$ erg/s, ``historical'' HMXBs in the LMC also agrees with
the value predicted from its global star formation rate. 
At lower luminosities, a deficit of sources is observed --  the
luminosity distribution seems 
to be somewhat flatter than the universal XLF (section
\ref{sec:hmxbcand}, Fig.\ref{fig:lfhmxb}).  
\item
We consider the impact of the ``propeller effect'' on the luminosity
distribution of HMXBs (section \ref{sec:propeller},
Fig.\ref{fig:proplum}) and demonstrate that it can explain
qualitatively the observed flattening of the XLF at low luminosities 
(Fig.\ref{fig:lfhmxb}). 
We note, however, that due to the relatively small number of HMXB
candidates in the observed part of the LMC and the lack of distinct signatures
of the ``propeller effect'' in the $\lg(L_X)\ga 33$ luminosity domain
(Fig.\ref{fig:proplum}),
it is premature to draw a definite conclusion about the relevance of
the ``propeller regime'' to the accretion onto a strongly magnetized
neutron star.
\item We found significant field-to-field variations in the number of
HMXBs, which appear to be uncorrelated with the star formation rates
inferred by FIR and $H_\alpha$ emission
(Fig.\ref{fig:halpha_src}). We suggest that these 
variations are related to different ages of the underlying stellar
population in different XMM-Newton fields. Using the existence of large
coeval stellar aggregates in the LMC we constrain the number of
HMXBs per unit stellar mass as a function of time elapsed since the
star formation event (section \ref{sec:age_effects},
Fig.\ref{fig:eta_hmxb}). Based 
on a simple ad hoc model, we obtain the theoretical dependence
$\eta_{HMXB}(\tau)$ and show that its normalization, as constrained
by the LMC data, is consistent with the calibration of the
$N_{HMXB}$--SFR relation, derived by \citet{grimm03}.
\end{enumerate}

\begin{acknowledgements}
This research has made use of data obtained through the High Energy 
Astrophysics Science Archive Research Center Online Service, provided 
by the NASA/Goddard Space Flight Center.
This publication has made use of data products from the Two Micron All
Sky Survey,  Guide Star Catalogue-II and USNO-B1.0 catalogue. 

PS would like to thank the
 Max-Planck-Institute for Astrophysics (Garching),
where a significant part of this project was done. 
PS also acknowledges partial support from the President of the
Russian Federation grant SS-2083.2003.2. 
\end{acknowledgements}

%\listofobjects

\end{document}